\documentclass[a4paper,fleqn,usenatbib,useAMS]{mnras}
\usepackage{graphicx}	
\usepackage{amsmath}	
\usepackage{amssymb}	
\usepackage{bm}		
\usepackage{pdflscape}

\usepackage[T1]{fontenc}
\usepackage{ae,aecompl}
\usepackage{longtable}

\title[Spatial Distribution of Globular Clusters in the Galaxy]{Spatial Distribution of Globular Clusters in the Galaxy}
\author[N. R. Arakelyan et al.]{
N. R. Arakelyan,$^{1,2}$\thanks{E-mail: n.rubenovna@mail.ru}
S. V. Pilipenko,$^{2}$
N. I. Libeskind$^{3}$
\\
$^{1}$Moscow Institute of Physics and Technology, Dolgoprudny, Institutskiy per., 9, Moscow Region, 141701, Russia \\
$^{2}$P.N. Lebedev Physical Institute of Russian Academy of Sciences, 84/32 Profsojuznaja Street, 117997 Moscow, Russia \\
$^{3}$Leibniz--Institut f\"{u}r Astrophysik, Potsdam, An der Sternwarte 16, D--14482 Potsdam, Germany
}

\date{Accepted XXX. Received YYY; in original form ZZZ}

\pubyear{2018}

\begin{document}
\label{firstpage}
\pagerange{\pageref{firstpage}--\pageref{lastpage}}
\maketitle

\begin{abstract}
The Milky Way's satellite galaxies and Globular Clusters (GCs) are known to exhibit an anisotropic spatial distribution. 
We examine in detail this anisotropy by the means of the inertia tensor. We estimate the statistical significance of the results by repeating this analysis for random catalogues which use the radial distribution of the real sample. Our method reproduces the well-known planar structure in the distribution of the satellite galaxies. We show that for GCs several anisotropic structures are observed. The GCs at small distances, $2<R<10$ kpc, show a structure coplanar with the Galactic plane. At smaller and larger distances the whole sample of GCs shows quite weak anisotropy. Nevertheless, at largest distances the orientation of the structure is close to that of the satellite galaxies, i.e. perpendicular to the Galactic plane. We estimate the probability of random realization for this structure of 1.7\%. The Bulge-Disk GCs show a clear disk-like structure lying within the galactic disk. The Old Halo GCs show two structures: a well pronounced polar elongated structure at $R<3$~kpc which is perpendicular to the galactic plane, and a less pronounced disk-like structure coplanar with the galactic disk at $6<R<20$~kpc. The Young Halo GCs do not show significant anisotropy.

\end{abstract}
\section{Introduction}

The presence of the cosmic web in the Universe indicates that the motions of matter are correlated on scales much larger than the size of a region from which a single galaxy collects its mass (e.g.,~\citet{1996Natur.380..603B,2005MNRAS.363..146L,2011MNRAS.411.1525L,2018MNRAS.473.1195L}). Thus, the cosmic web should be connected with the distribution of matter inside galaxies. This effect is seen as a presence of anisotropic distribution of satellites around their host galaxies (see, e.g., \citet{2015MNRAS.452.1052L}). The Milky Way has a population of satellites that forms a disk-like structure almost perpendicular to the disk of our Galaxy (e.g.~\citet{2005A&A...431..517K, 2008ApJ...680..287M}). Historically, this was first identified for four satellite galaxies \citep{1976MNRAS.174..695L, 1983IAUS..100...89L, 1982Obs...102..202L, 1995MNRAS.275..429L}, but over time this number reached eleven \citep{2008ApJ...680..287M, 2007MNRAS.374.1125M}, more than twenty \citep{2009MNRAS.394.2223M}, and then this was confirmed for 14 newly discovered satellites in the Southern hemisphere \citep{2015MNRAS.453.1047P}. The Andromeda galaxy also shows a ``vast, thin, co-rotating'' plane of satellite galaxies \citep{2013Natur.493...62I,2013ApJ...766..120C}, although this stricture is not polar, like in the Milky Way. SDSS galaxies show some lopsidedness in the distribution of satellites, directed towards a closest other massive galaxy \citep{2016ApJ...830..121L} which is also expected from cosmological N-body simulations \citep{2017ApJ...850..132P}.

Besides satellite galaxies,  galaxy halos are also populated by Globular Clusters (GCs). The GCs have ages up to 13 Gyr, so they may represent fossils of the earliest stages of galaxy evolution. Some of the GCs may originate from satellite galaxies that have been themselves been accreted by the Galaxy \citep{1998ApJ...501..554C,2002ApJ...567..853C,2002MNRAS.333..383B,2004MNRAS.355..504M,2005ApJ...623..650K,2006ARA&A..44..193B}. GCs are usually divided into three classical types: young halo (YH), old halo (OH), bulge disk (BD) \citep{1978ApJ...225..357S,1985ApJ...293..424Z,2005MNRAS.360..631M}. It is expected that most of the YH GCs and around 15\%--17\% of the OH GCs may have extragalactic origin \citep{2004MNRAS.355..504M}.

Since the currently existing satellites of the Milky Way form a polar plane around the Milky Way, it is attractive to search for the same anisotropy in the distribution of GCs. This has been done in several papers \citep{2002Sci...297..578Y,2005A&A...431..517K,2009arXiv0906.5370F,2012ApJ...744...57K,2012MNRAS.423.1109P}. In particular, \citet{2012ApJ...744...57K} analysed the distribution of the YH and OH GCs. They perform the best-fitting of a plane in the distribution of GCs, taking into account the uncertainties in the GCs distance measurements to estimate the plane robustness. They found that YH GCs form a plane with orientation very close to that of the satellites. By analysing the samples located at galactocentric distances $>10$, $>20$ and $>30$~kpc they have also found that the GCs distribution becomes more confined to a preferred direction at larger distances. In contrast, OH GCs do not show an evidence of a planar structure.
\citet{2012MNRAS.423.1109P} confirm the preferred orientation of YH GCs using a different method of the search for a plane, similar to that used in \citet{2010A&A...523A..32K}. They also extended the analysis by adding 14 stellar streams, dynamically stripped from accreted  satellites: they found that 7 of these streams lie within the Milky Way's polar structure.

The aim of this paper is to bring more detail in the examination of the GC anisotropy. First of all, previous works assumed a planar distribution and tried to define the plane that best matches the distribution of GCs. Instead, we use the inertia tensor and perform a ``blind search'' for anisotropy.
Second, the system of GCs undergoes changes due to the interaction between the GCs and the Galactic disk (e.g.~\citet{1997ApJ...474..223G}), so one might expect that the GCs which have been accreted more recently and/or are located further from the Milky Way center, to conserve a memory on their anisotropic accretion. In particular, GCs with galactocentric distances $>10$~kpc are expected to be long-lived, with lifetimes exceeding the Hubble time. To examine this issue we not only divide GCs into the three classical groups following \citet{2005MNRAS.360..631M}, but also give all characteristics of the anisotropy as a function of distance from the Galactic center. Previous works have focused mainly on the distant GCs with $R>10$~kpc while we analyse anisotropy at all distances.

This paper is organized as follows: in Section 2 we describe the method of measuring the anisotropy. In Section 3 we present the results of measurements for the whole sample of GCs. In Section 4 we analyse the three classical types of GCs. In Section 5 we present discussion of our findings and conclusions.

\section{Anisotropy Measurement}
\begin{figure*}
 \includegraphics[width=0.49\textwidth]{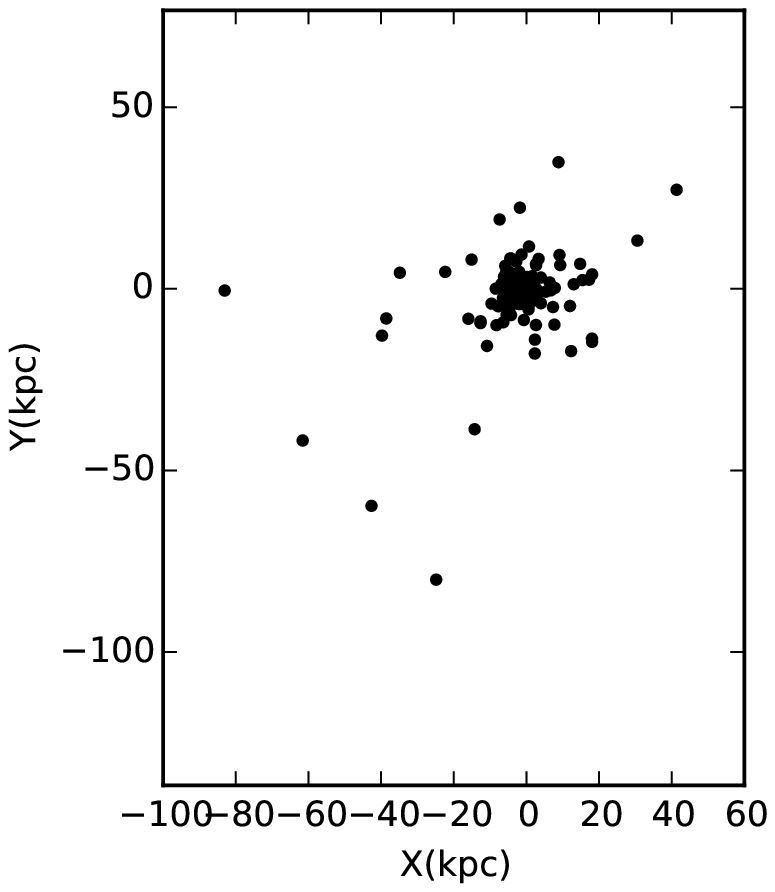} 
 \includegraphics[width=0.49\textwidth]{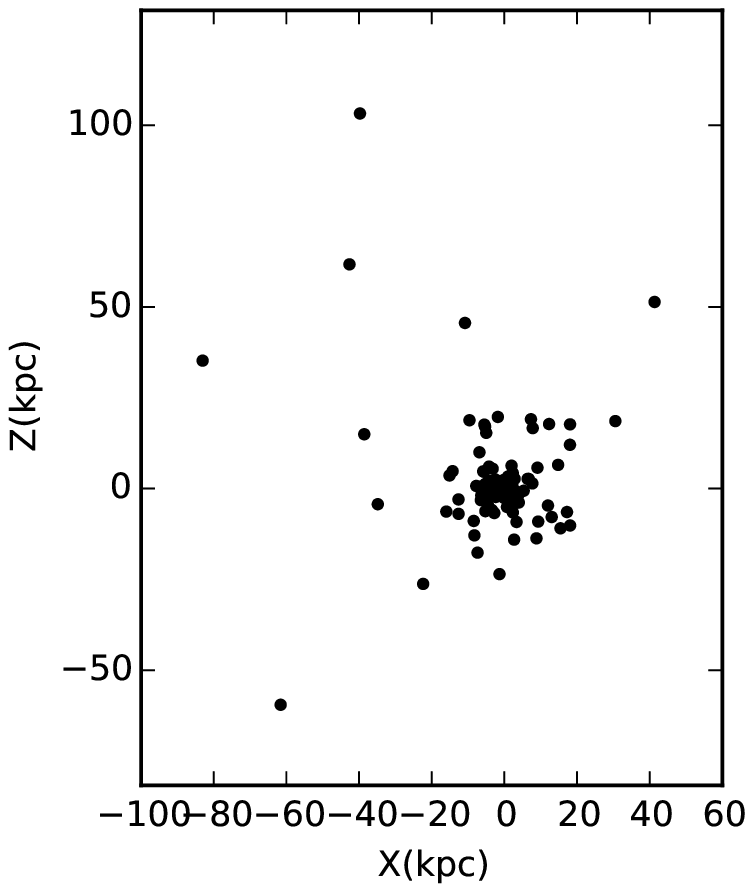} 
 \caption{Distribution of GCs. XYZ -- Cartesian coordinates relative to the center of the Galaxy. Z is aligned with the Galactic pole and coordinates of the Sun are (-8.34, 0, 0)}
 \label{fig:1}
\end{figure*}

\begin{figure*}
 \includegraphics[width=0.49\textwidth]{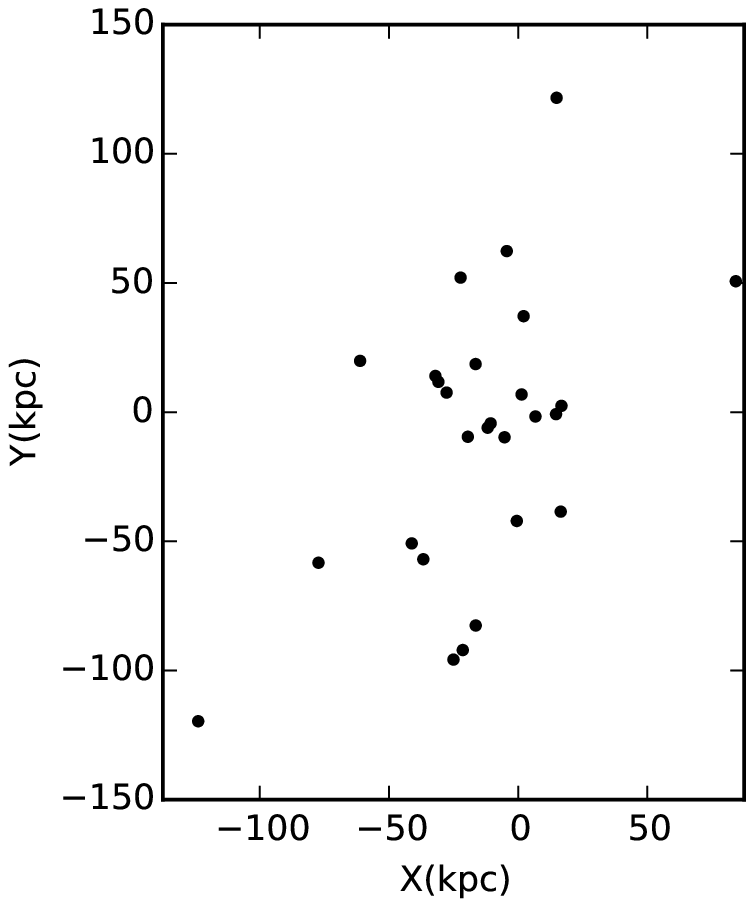} 
 \includegraphics[width=0.49\textwidth]{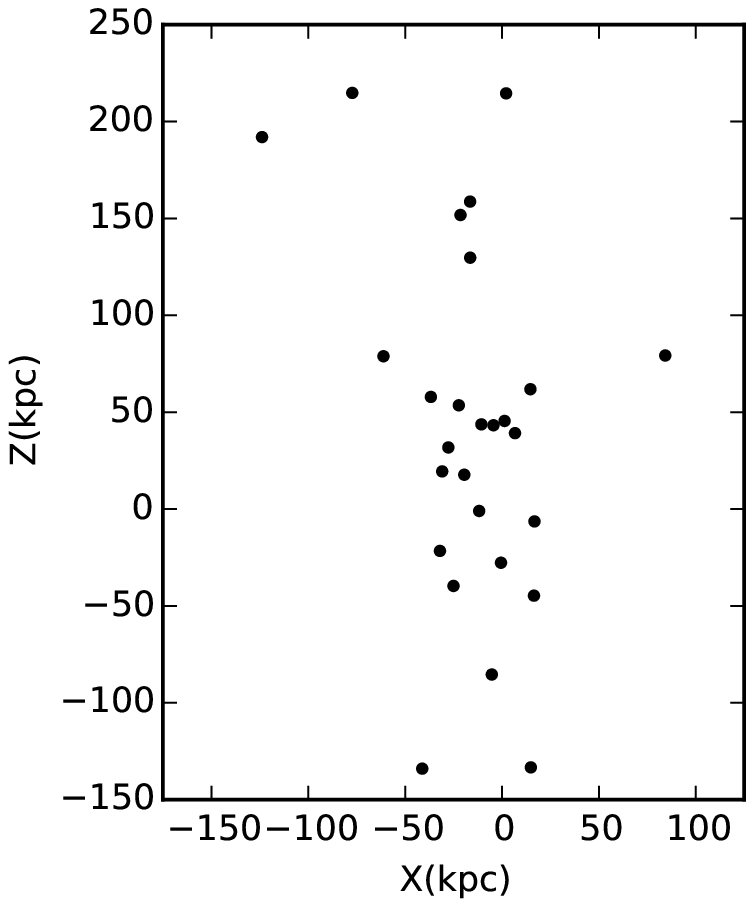} 
 \caption{Distribution of satellite galaxies. Coordinate axes as in Fig.1}
 \label{fig:2}
\end{figure*}

Our Galaxy contains at least 157 GCs \citep{2013ApJ...772...82H} \footnote{\url{http://physwww.mcmaster.ca/~harris/Databases.html}}, and 27 satellite galaxies (the number of satellite galaxies of the local group is 53) \citep{2012AJ....144....4M}. The catalogues we use are presented in Tables~\ref{tab:1} and \ref{tab:2} (full version is available online). In Figure~\ref{fig:1}, we show the GCs distribution in galactocentric distances in kpc. The distribution is represented in x--y and x--z planes of the Cartesian coordinates, relative to the center of the Galaxy. The plane x--y corresponds to the disk of the Galaxy. In the Figure, it is easy to see that the halo of the Galaxy extends approximately 100 kpc. The distribution of satellite galaxies can be seen in Figure~\ref{fig:2}. Here, too, the galactocentric distances are shown in the x--y and x--z planes. The anisotropy in the distribution of satellites is seen by a naked eye in the right panel of Figure~\ref{fig:2}, while the distribution of GCs does not show such a clear effect.

\begin{table*}
 \caption{Catalogue of Galactic GCs coordinates and types compiled from \citet{2013ApJ...772...82H} and \citet{2005MNRAS.360..631M} (full version available online).}
 \label{tab:1} 
 \medskip
 \begin{tabular}{|l|c|c|c|c|}
  \hline
  \multicolumn{1}{|c|}{Name$^{1)}$} &  l$^{2)}$ & b$^{3)}$   & R-Sun$^{4)}$   & Type$^{5)}$ \\
  \hline 
  NGC 104   & 305.89  & -44.89   & 4.5   & BD \\
  \hline  
  NGC 288   & 152.30   & -89.38   & 8.9   & OH \\
  \hline 
  NGC 362   & 301.53   & -46.25   & 8.6   & YH \\
  \hline
  Whiting 1 & 161.22   & -60.76   & 30.1  & UN \\
  \hline
  NGC 1261  & 270.54   & -52.12   & 16.3  & YH \\
  \hline   
  \end{tabular}
  \begin{tabular}{rp{150mm}}
 $^{1)}$  &  Name  \\
 $^{2)}$  &  Galactic longitude (deg) \\
 $^{3)}$  &  Galactic latitude (deg)   \\
 $^{4)}$  &  Distance from the Sun (kpc)  \\
 $^{5)}$  &  Type, according to \citet{2005MNRAS.360..631M} \\
\end{tabular}
\end{table*}

\begin{table*}
 \caption{Catalogue of Milky Way satellite galaxies extracted from \citep{2012AJ....144....4M} (full version available online).}
 \label{tab:2} 
 \medskip
 \begin{tabular}{|l|c|c|c|}
  \hline
  \multicolumn{1}{|c|}{Name$^{1)}$} &  l$^{2)}$ &  b$^{3)}$ & R-Sun$^{4)}$  \\
  \hline 
Canis Major   & 240.0   & -8.0    & 7  \\
  \hline   
  Sagittarius dSph   & 5.6    & -14.2    & 26 \\
  \hline
  Segue (I)          & 220.5  & +50.4    & 23 \\
  \hline
  Ursa Major II      & 152.5  & +37.4    & 32 \\
  \hline
  Bootes II          & 353.7  & +68.9    & 42 \\
  \hline  
  \end{tabular}
  \begin{tabular}{rp{150mm}}
  $^{1)}$  &  Name  \\
  $^{2)}$  &  Galactic longitude (deg) \\
  $^{3)}$  &  Galactic latitude (deg)  \\
  $^{4)}$  &  Distance from the Sun (kpc)  \\
 \end{tabular}
\end{table*}

To test the anisotropy of distributions quantitatively, we use two different tensors: the inertia tensor and the reduced tensor, which are constructed as follows:

\begin{equation}
\label{form:1} 
 S_{ij}=\frac{1}{N}\sum_{k=1}^Nx_i^kx_j^k,
 \end{equation} 
 
 \begin{equation}
\label{form:2} 
 J_{ij}=\frac{1}{N}\sum_{k=1}^N\frac{x_i^kx_j^k}{R_k^2},
 \end{equation}  
where  $S$ is the inertia tensor, $J$ is the reduced tensor, $N$ is the number of objects, $x_i^k$ is the distance from $k$-th object to the center of the Galaxy along $i$-th coordinate axis, $R_k^2 = x_k^2 + y_k^2 + z_k^2$, $R$ is the distance to each $k$-th particle. The three eigenvalues of the inertia tensor (a,b and c) are sorted in increasing order such that a>b>c. The degree of the anisotropy is characterized by the ratios of the eigenvalues, $c/a$ and $b/a$, both of which approach to 1 in case of isotropic distribution. The eigenvectors of the tensor give us the directions of anisotropy.

\begin{figure*}
 \includegraphics[scale=0.8]{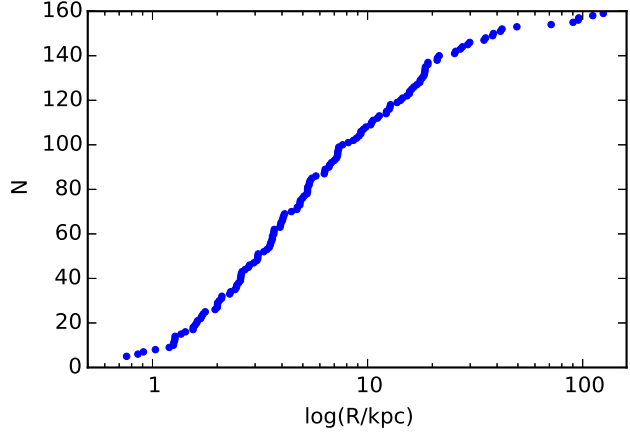}
 \caption{Distribution of GCs by distance. N -- number of objects; logR -- distance from the center of the Galaxy.}
 \label{fig:4}
\end{figure*}

Since the terms of the sum in equation (\ref{form:1}) are proportional to the square of distance, the inertia tensor is very sensitive to the presence of objects with large distances. The reduced tensor (\ref{form:2}), in contrast, does not take distances into account. Since the GCs are very concentrated towards the Galactic center (see Figure~\ref{fig:4}), the central part of the Galaxy gives the highest contribution to the reduced tensor. We show the impact of these effects below.

\subsection{Statistical significance of anisotropy}
\label{sec:metod I} 

\begin{figure*}
 \includegraphics[width=0.49\textwidth]{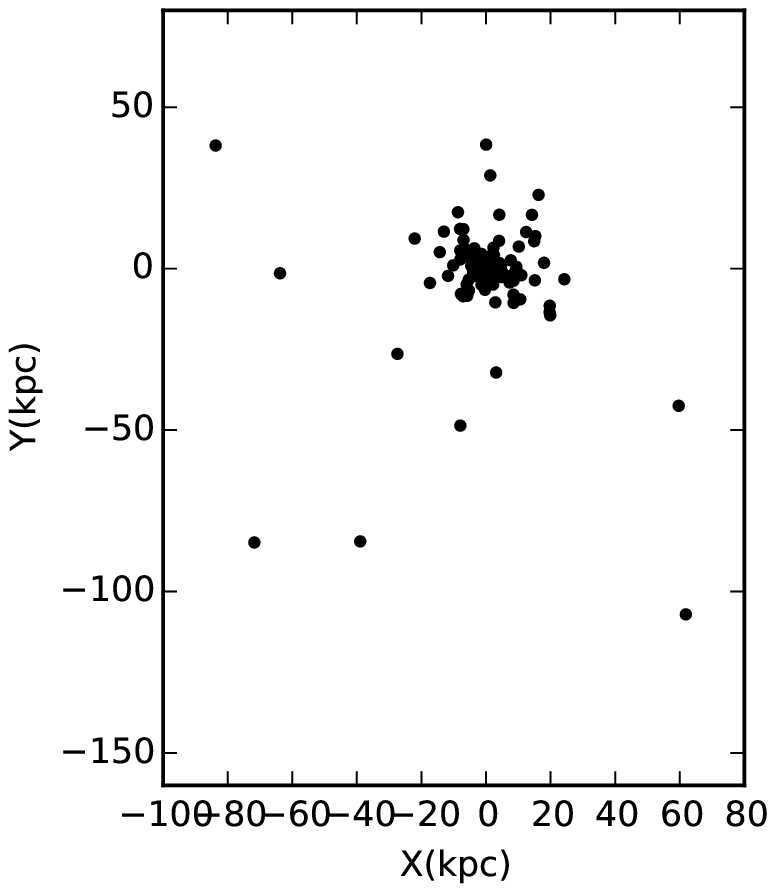} 
 \caption{One random samples generation. Coordinate axes as the same as Fig.1 }
 \label{fig:3} 
\end{figure*}

To check the statistical significance of the found parameters of the system of GCs we generate 10,000 random samples with the same radial distribution and number of objects as the data, and measure median value and RMS of the ratio of  tensors eigenvalues. We call the anisotropy statistically significant if the ratio of tensor eigenvalues for a real catalogues differs from the median of the random samples by more than $3\sigma$. The random samples are constructed by fixing the distances ($R$) from the real sample and randomizing the angular coordinates.
In Figure~\ref{fig:3} we show the distribution of GCs for one sample with random coordinates of objects.

\section{Anisotropy as a Function of Distance}

Figures from 4 to 10 show the results of anisotropy measurements for GCs and satellite galaxies (SGs) with the help of the inertia tensor and the reduced tensor. 
In the figures that show $c/a$ and $b/a$ as a function of $R$ the real object distributions are represented by dots, the solid line represents the median result for 10,000 random samples, and the dashed lines represent the median $\pm3$\,$\sigma$. ``Angle'' on these figures is measured between the normal to the plane of the Galaxy and the minor (green triangles) or the major (blue dots) axes of the distribution. 
Since the inertia tensor is sensitive to the most distant objects distribution, the results in the top row of plots for a given value of $R$ may be interpreted as describing anisotropy close to $R$, while for the reduced tensor as describing the cumulative properties at $<R$.

\begin{figure*}
 \includegraphics[scale=0.8]{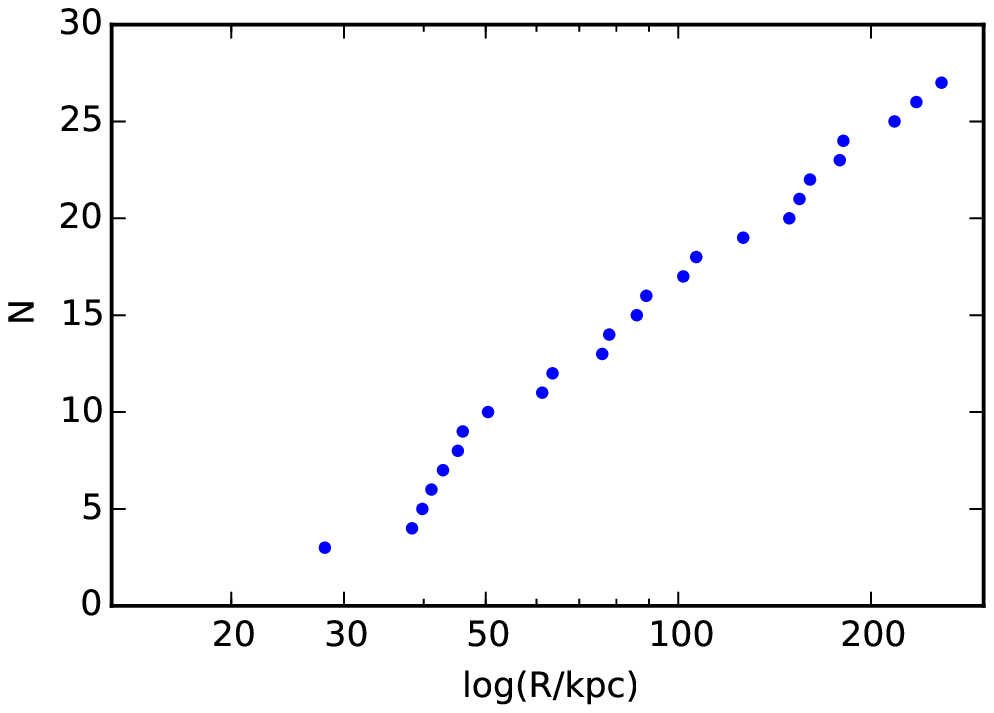}
 \caption{Distribution of satellite galaxies by distance. N -- number of objects; logR -- distance from the center of the Galaxy.}
 \label{fig:8}
\end{figure*}  

\begin{figure*}
 \includegraphics[width=0.325\textwidth]{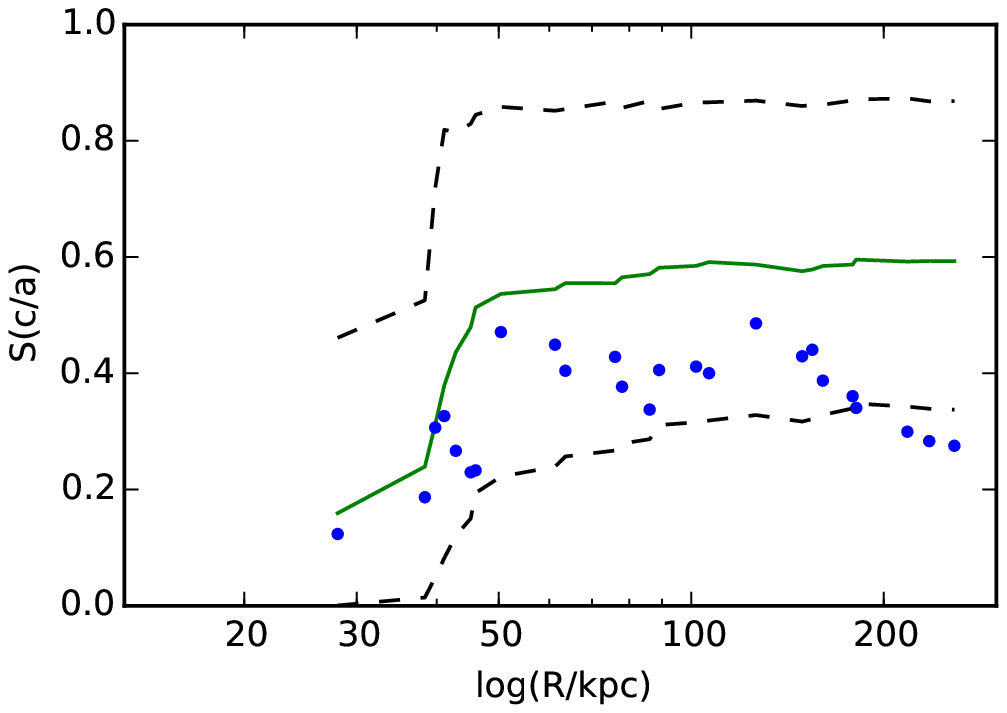}
 \includegraphics[width=0.325\textwidth]{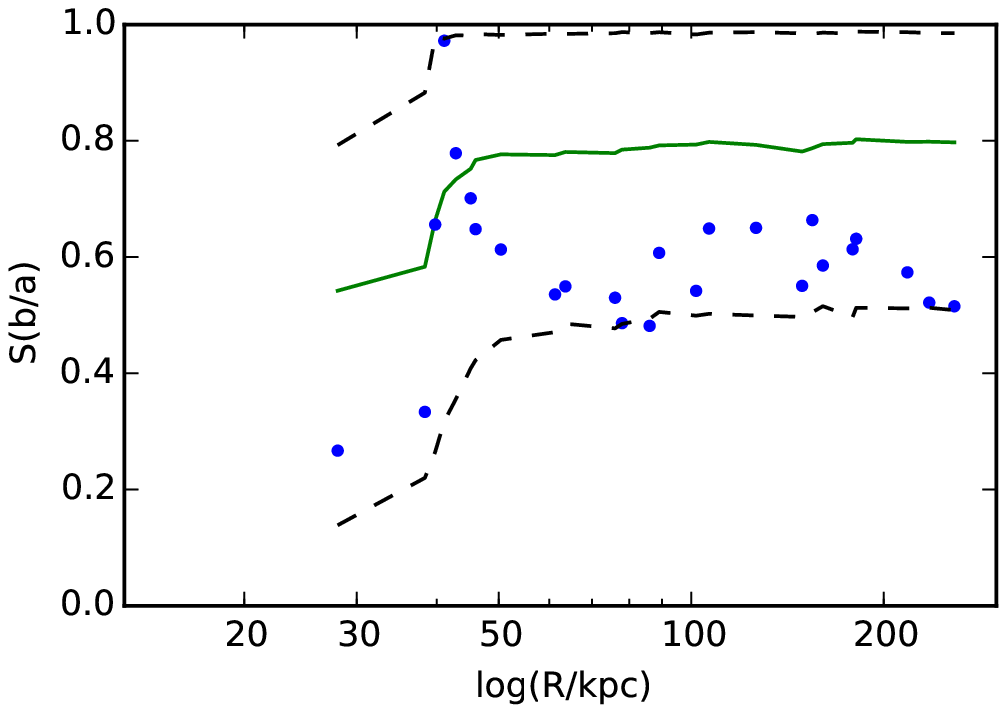}
 \includegraphics[width=0.325\textwidth]{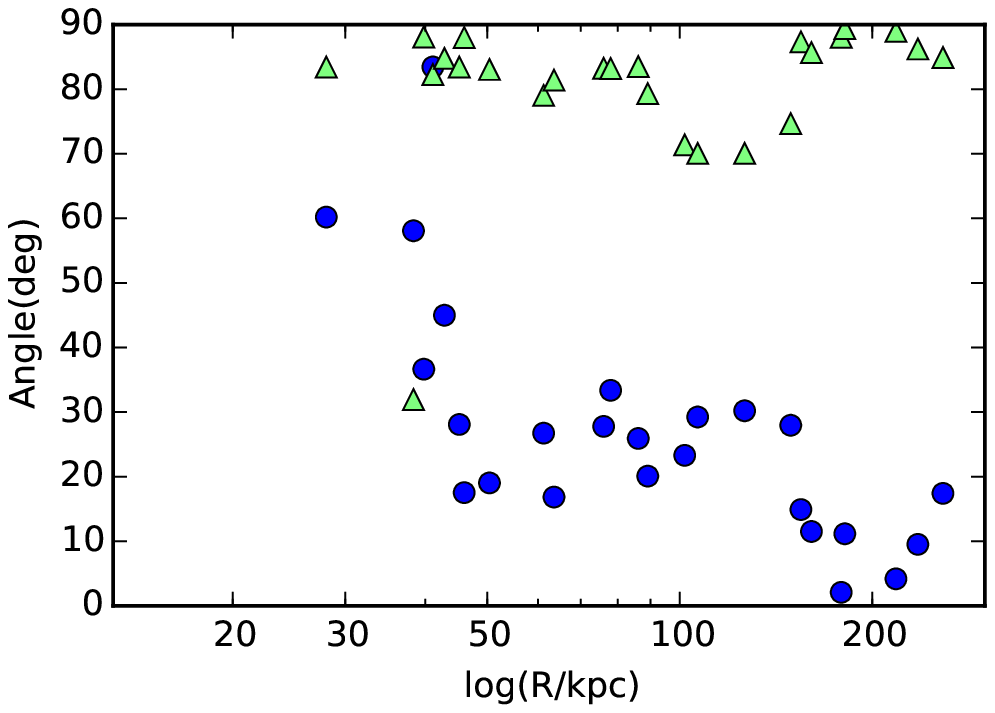} 
 
 \includegraphics[width=0.325\textwidth]{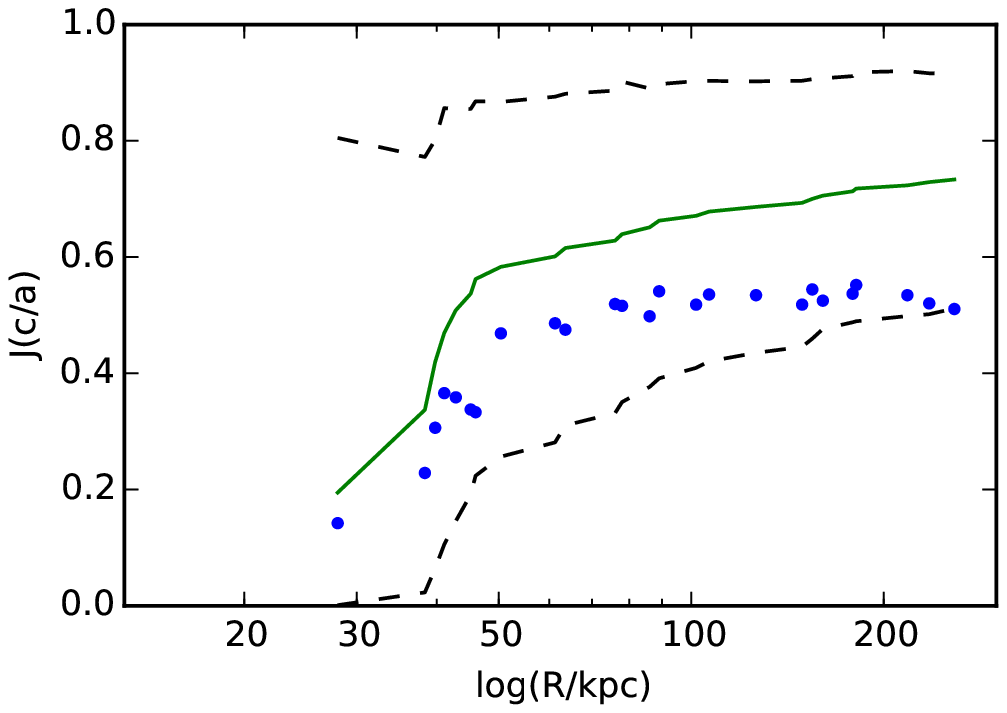}
 \includegraphics[width=0.325\textwidth]{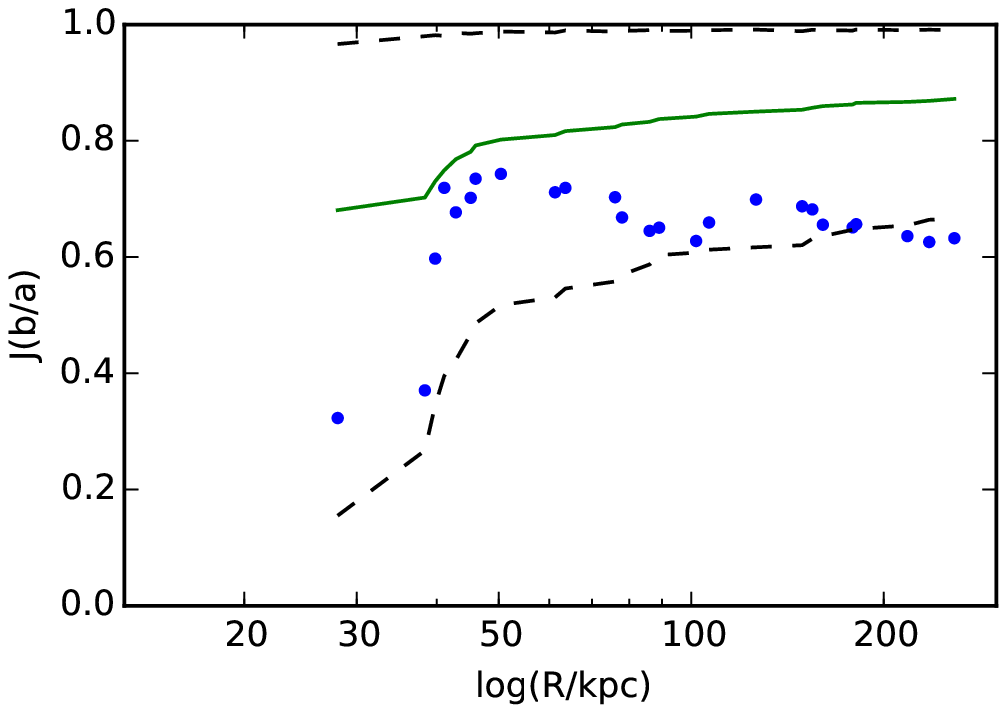}
 \includegraphics[width=0.325\textwidth]{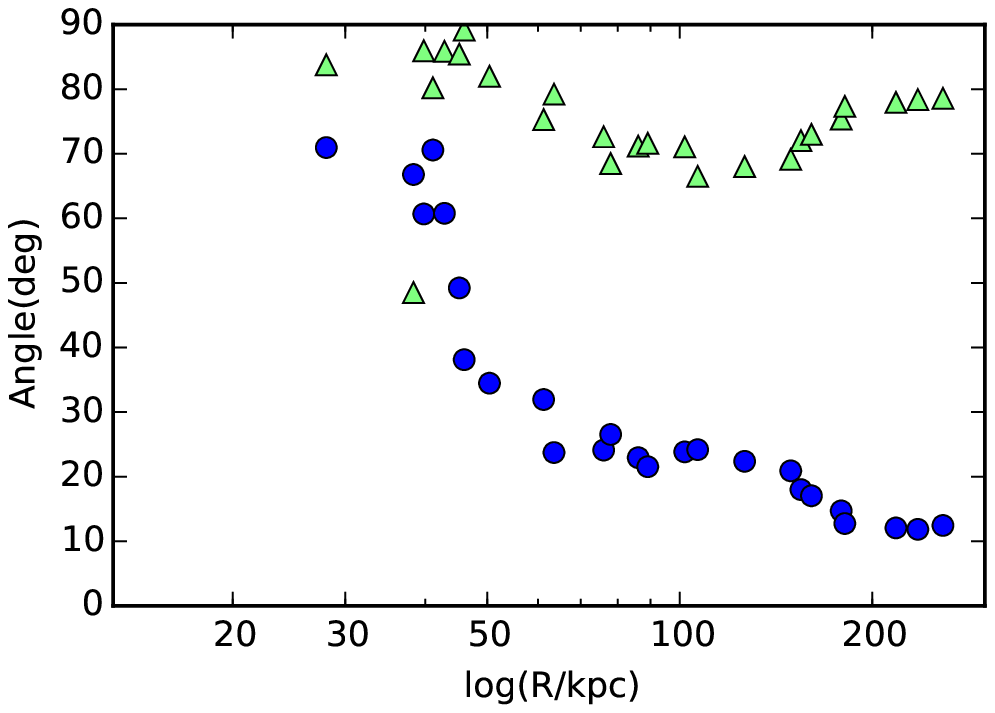}
 \caption{ The Anisotropy of 27 satellite galaxies quantified by the inertia tensor (equation (1), top row) and the reduced inertia tensor (equation (2), bottom row). The left column shows the distribution of $c/a$ as function of satellite galactocentric distance. The middle column shows the distribution of $b/a$ as a function of galactocentric distance. Each blue dot represents the cumulative eigenvalue ratio of these tensors computed for all galaxies {\it interior} to that position. The solid green line represents the median eigenvalue ratios for 10,000 random samples that maintain the same radial distribution as the data, but whose polar and azimuthal angle has been randomised. The dashed lines represent the $\pm3\sigma$ of such random distributions. The right column shows the angle, measured in degrees, subtended between the Milky Way$'$s galactic pole and the major (blue dots) and the minor (green triangles) axis of the two inertia tensors. Green triangles close to 90 deg indicate a polar plane.}
 \label{fig:9}
\end{figure*}

\begin{figure*}
 \includegraphics[width=0.325\textwidth]{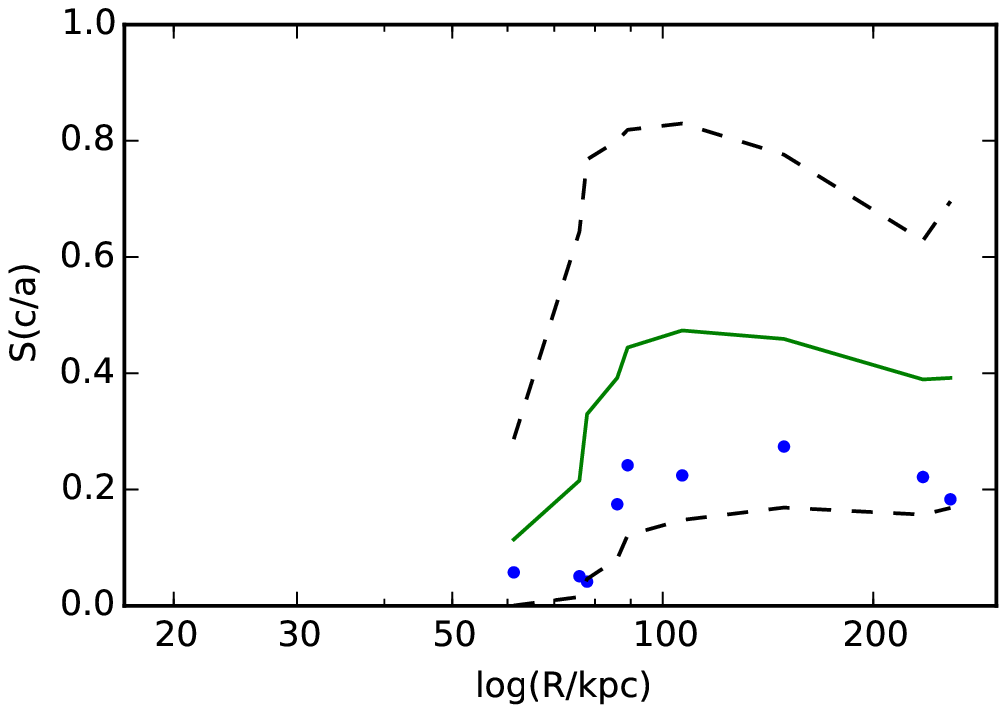}
 \includegraphics[width=0.325\textwidth]{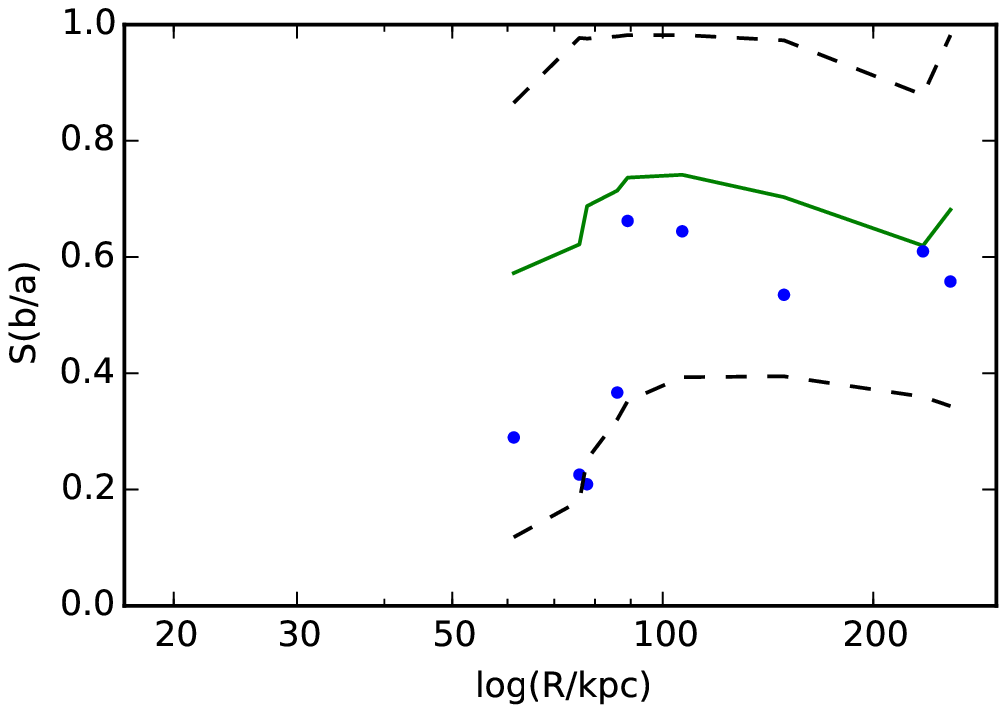}
 \includegraphics[width=0.325\textwidth]{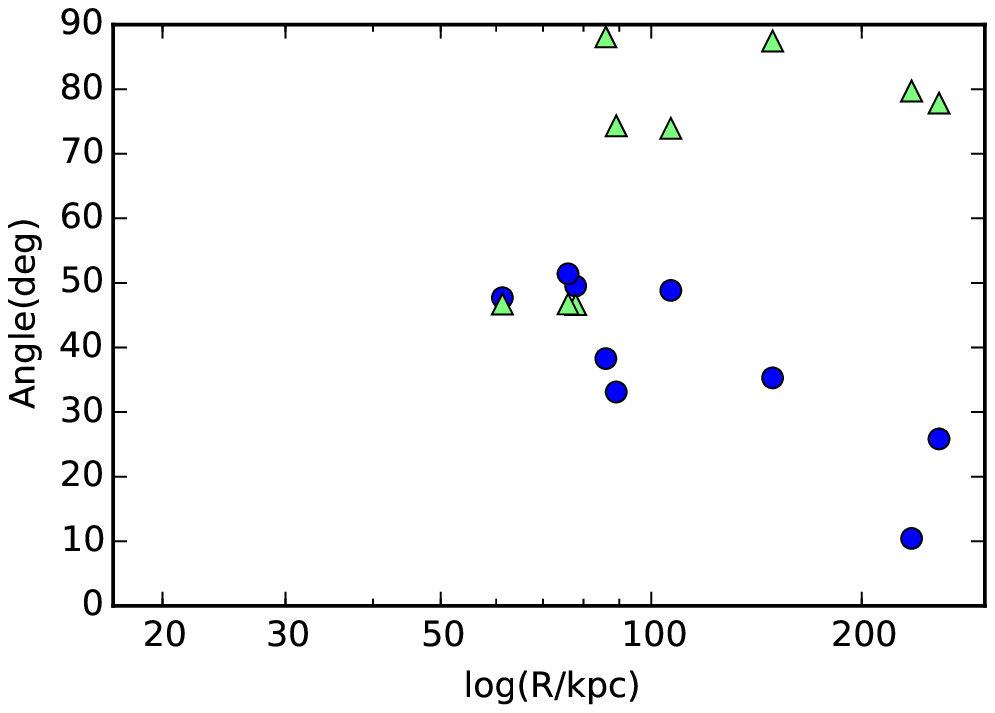} 
 
 \includegraphics[width=0.325\textwidth]{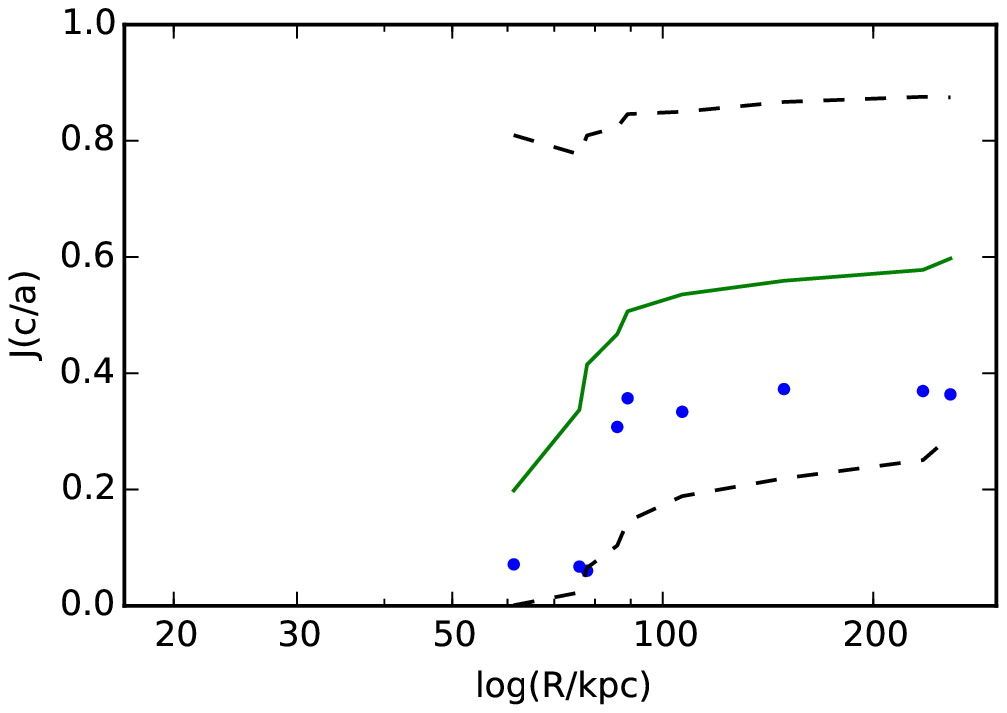}
 \includegraphics[width=0.325\textwidth]{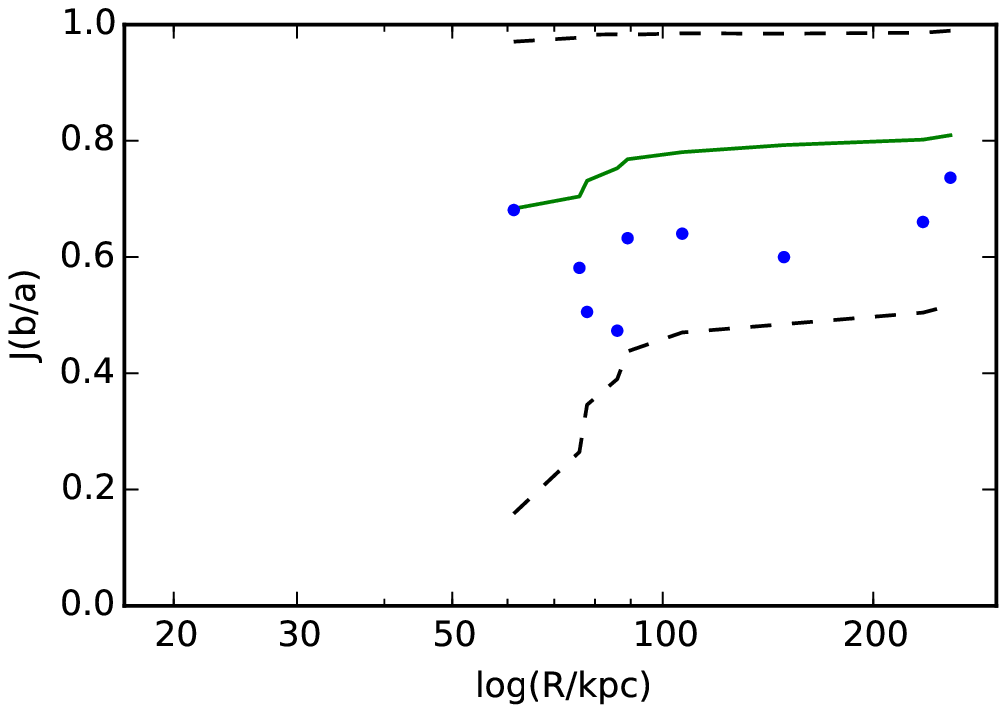}
 \includegraphics[width=0.325\textwidth]{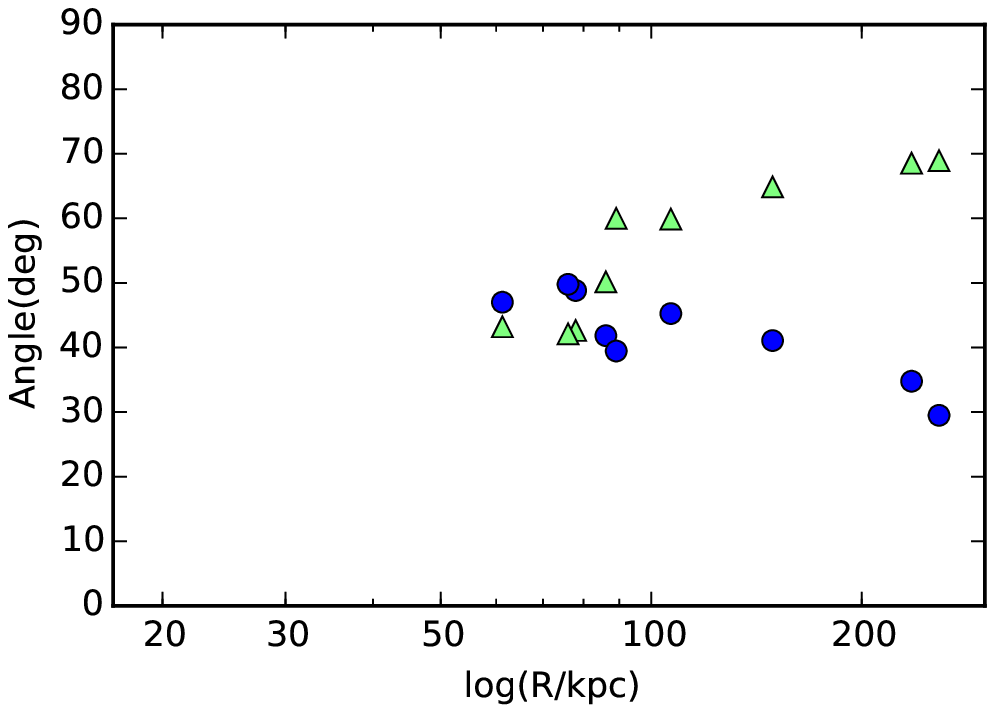}
 \caption{Same as Fig.5, but for 11 satellite galaxies. }
 \label{fig:10}
\end{figure*}

\begin{figure*}
 \includegraphics[width=0.325\textwidth]{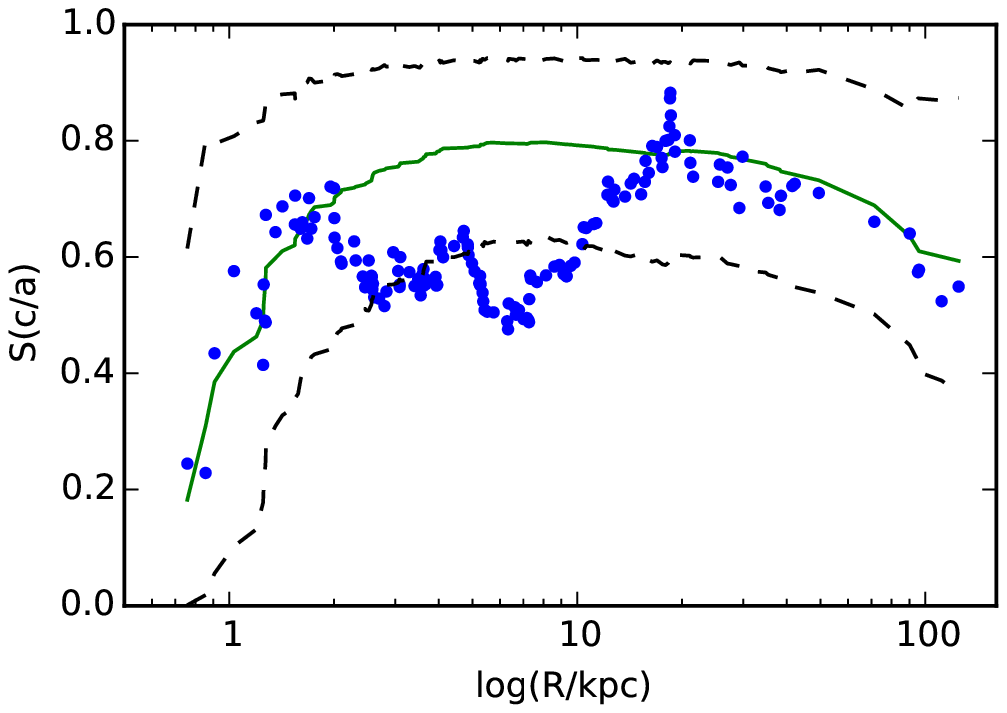}
 \includegraphics[width=0.325\textwidth]{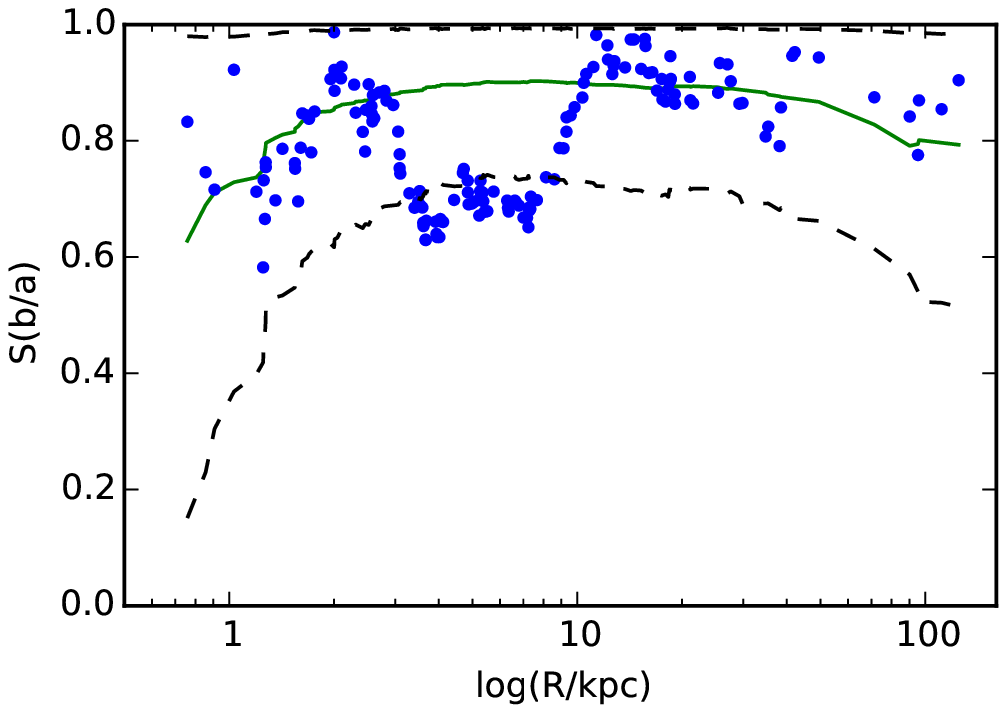}
 \includegraphics[width=0.325\textwidth]{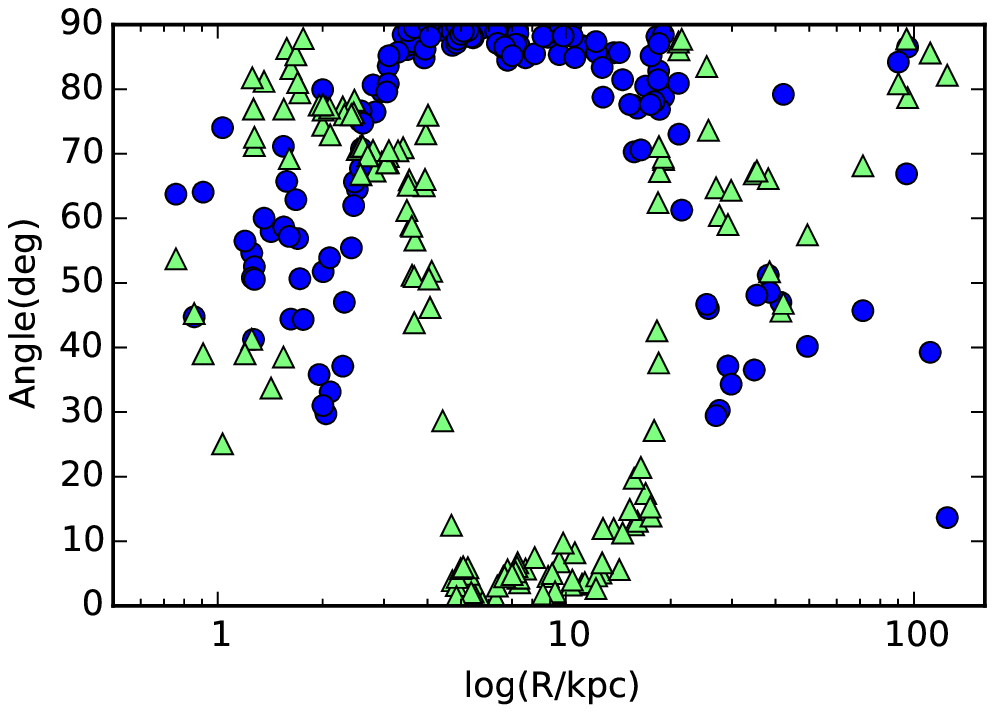} 
 
 \includegraphics[width=0.325\textwidth]{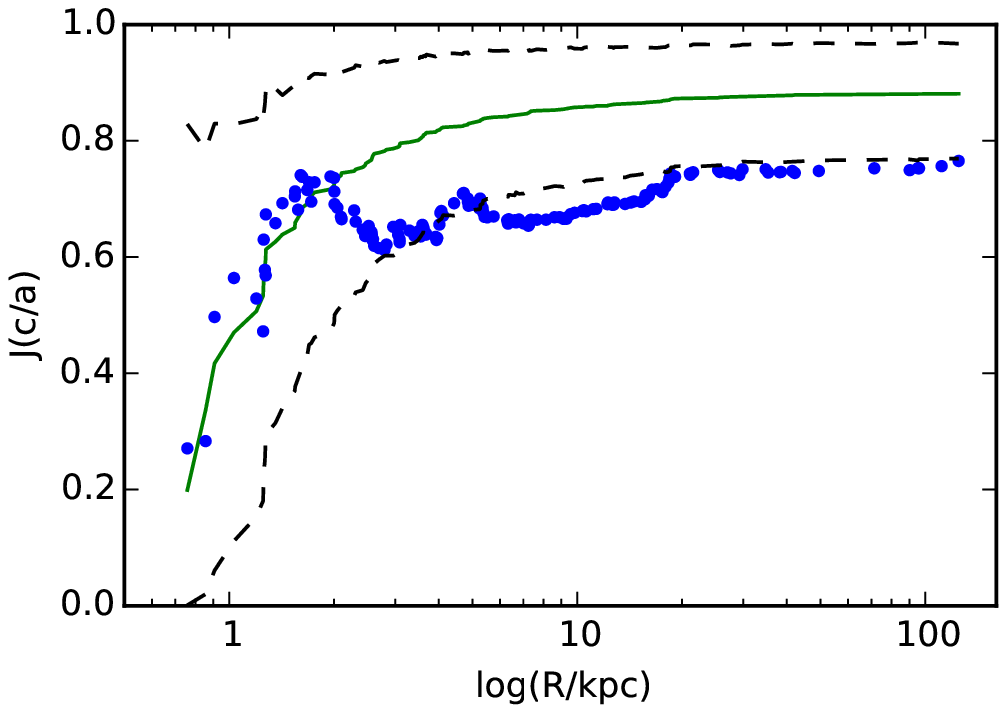}
 \includegraphics[width=0.325\textwidth]{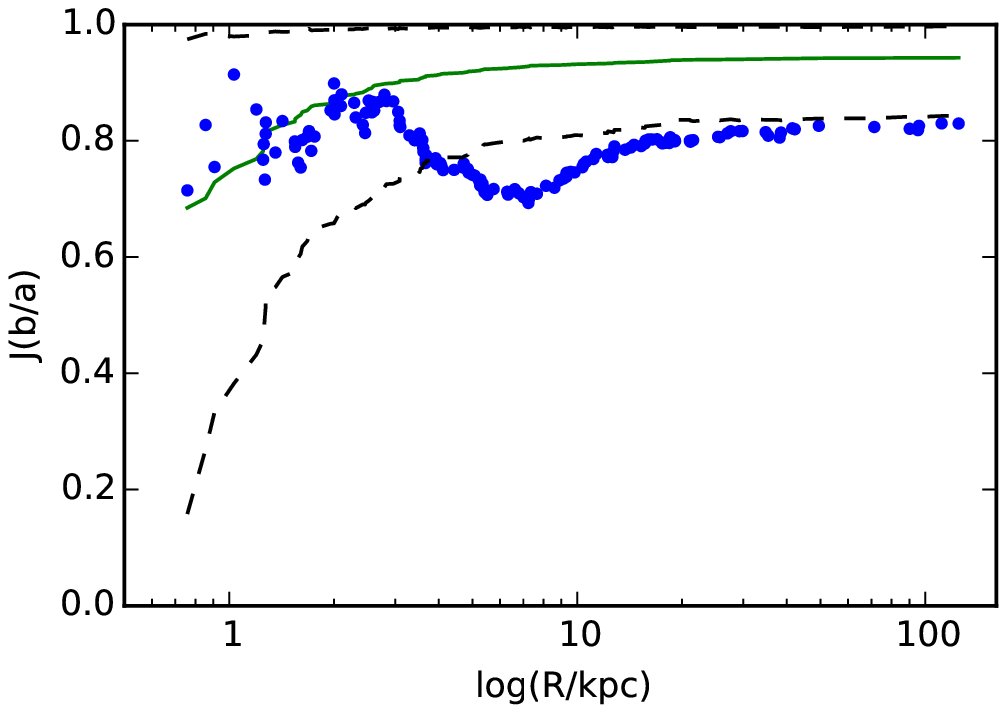}
 \includegraphics[width=0.325\textwidth]{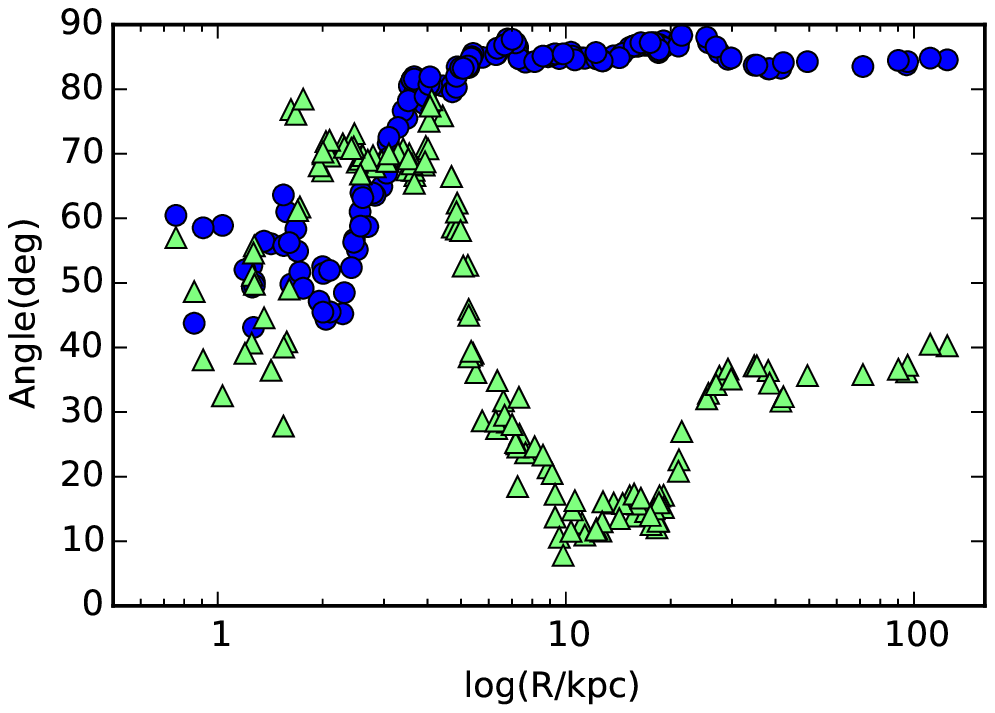}
 \caption{Same as Fig.5, but for GCs.  }
 \label{fig:5}
\end{figure*}

\begin{figure*}
 \includegraphics[width=0.325\textwidth]{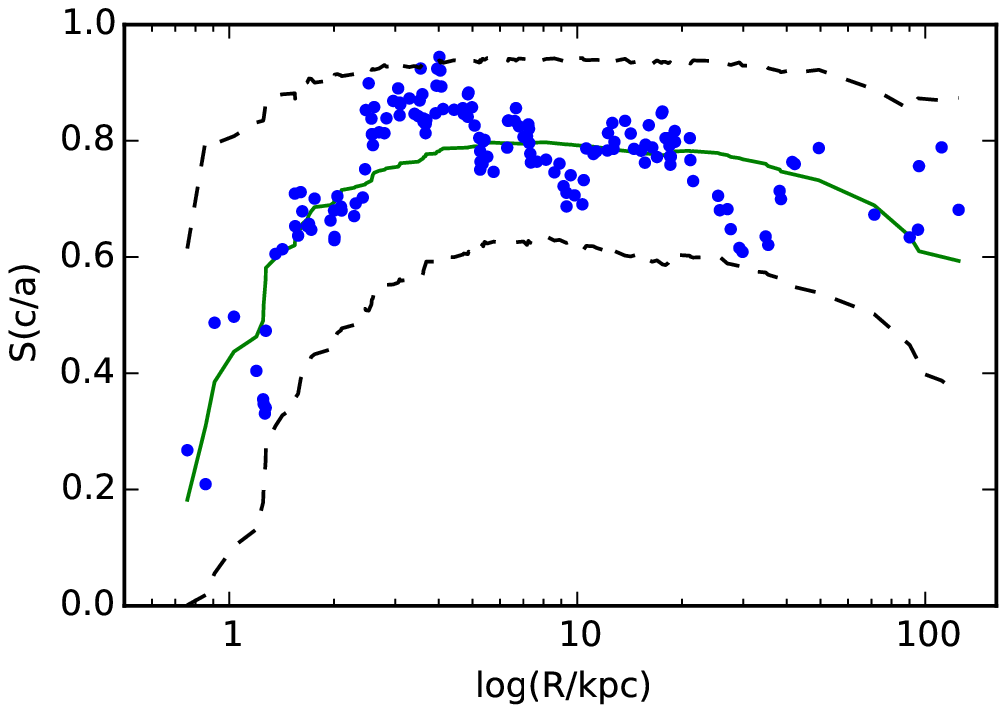}
 \includegraphics[width=0.325\textwidth]{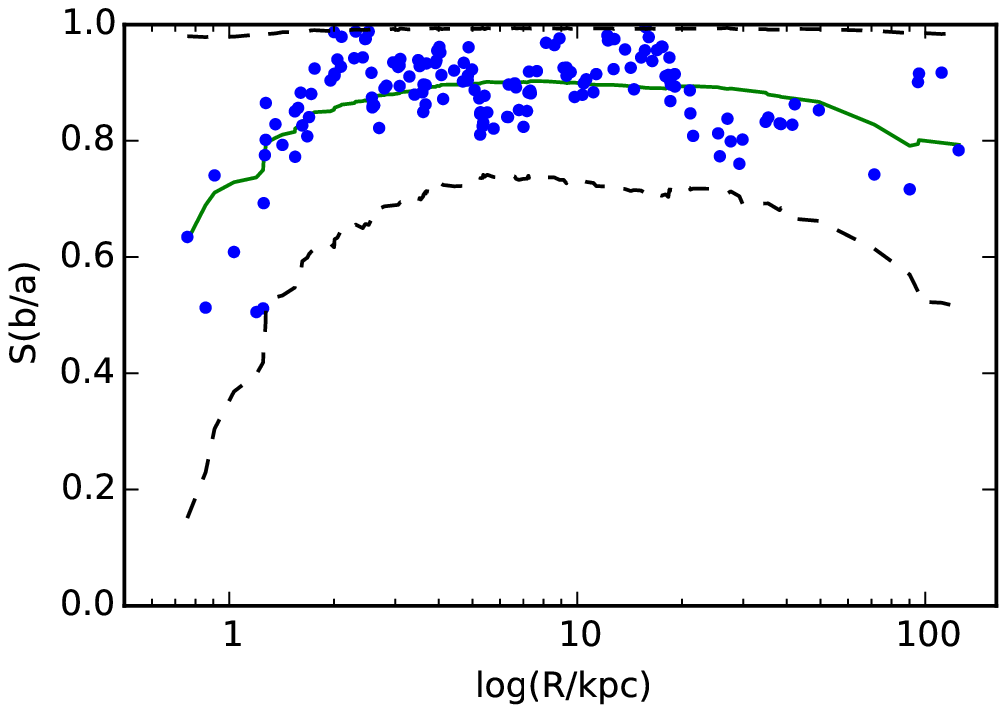}
 \includegraphics[width=0.325\textwidth]{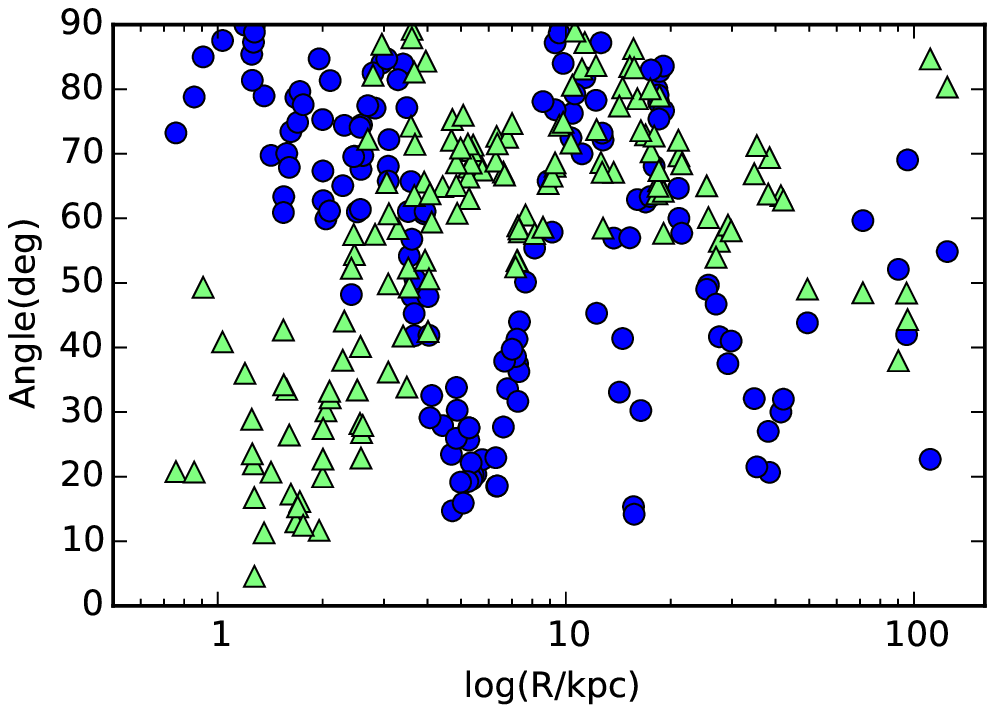}

 \includegraphics[width=0.325\textwidth]{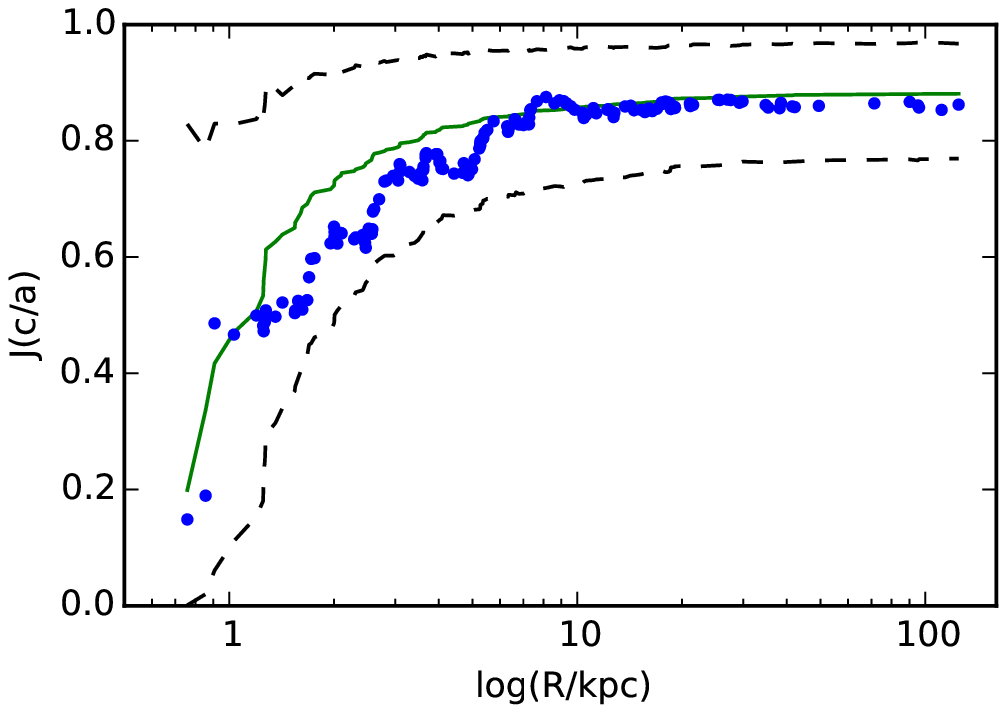}
 \includegraphics[width=0.325\textwidth]{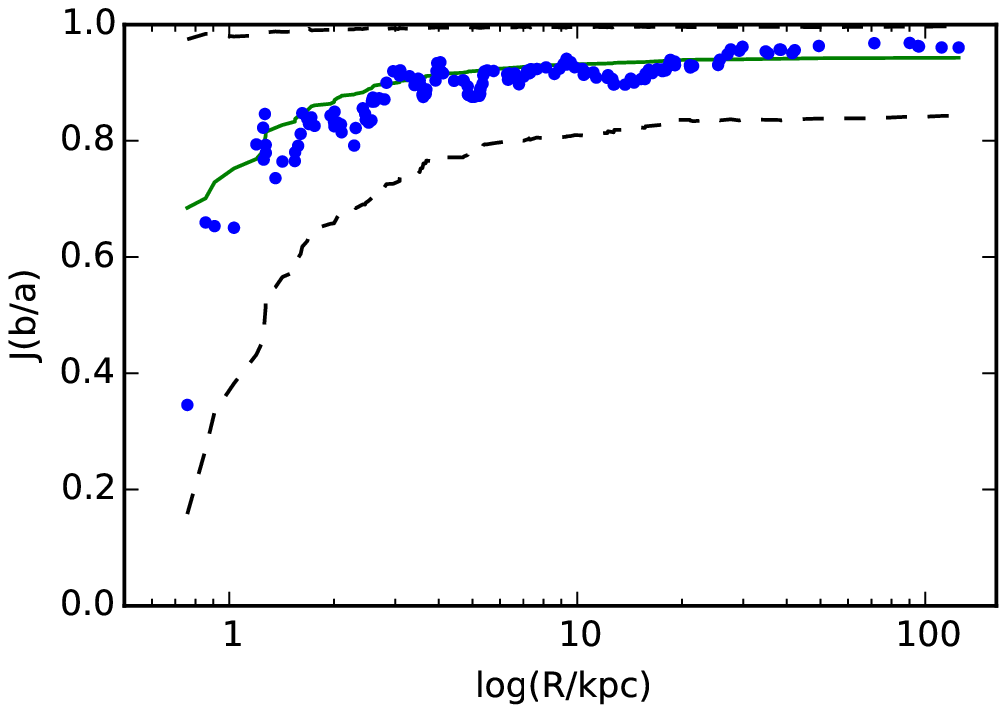}
 \includegraphics[width=0.325\textwidth]{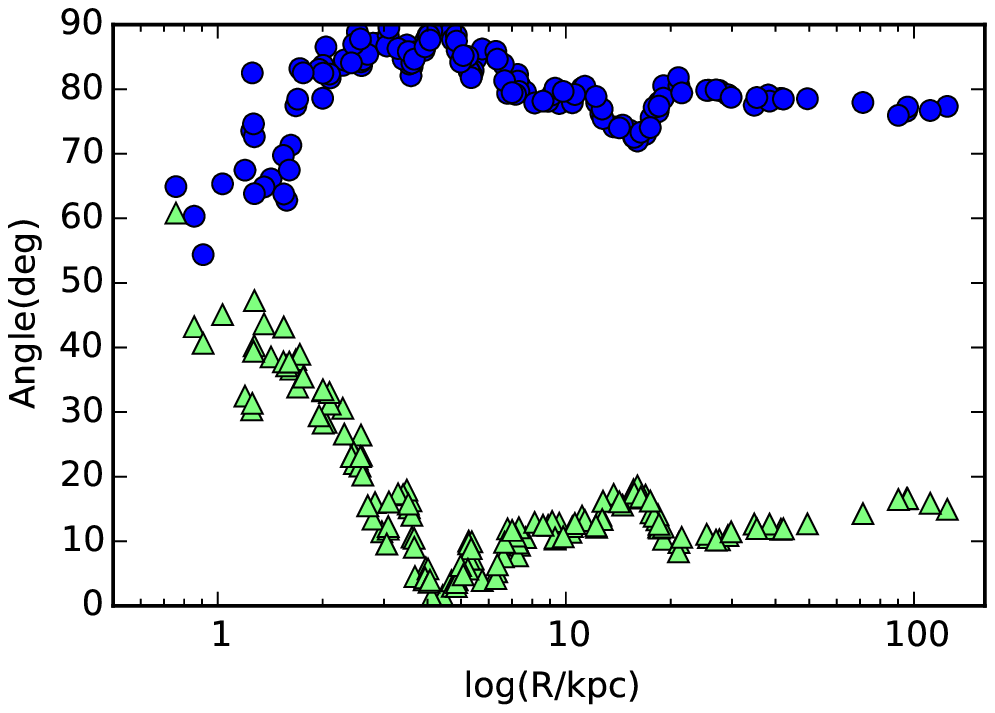}
 \caption{ Same as Fig.5, but for one random sample generation for GCs.}
 \label{fig:6}
\end{figure*}

\begin{figure*}
 \includegraphics[width=0.325\textwidth]{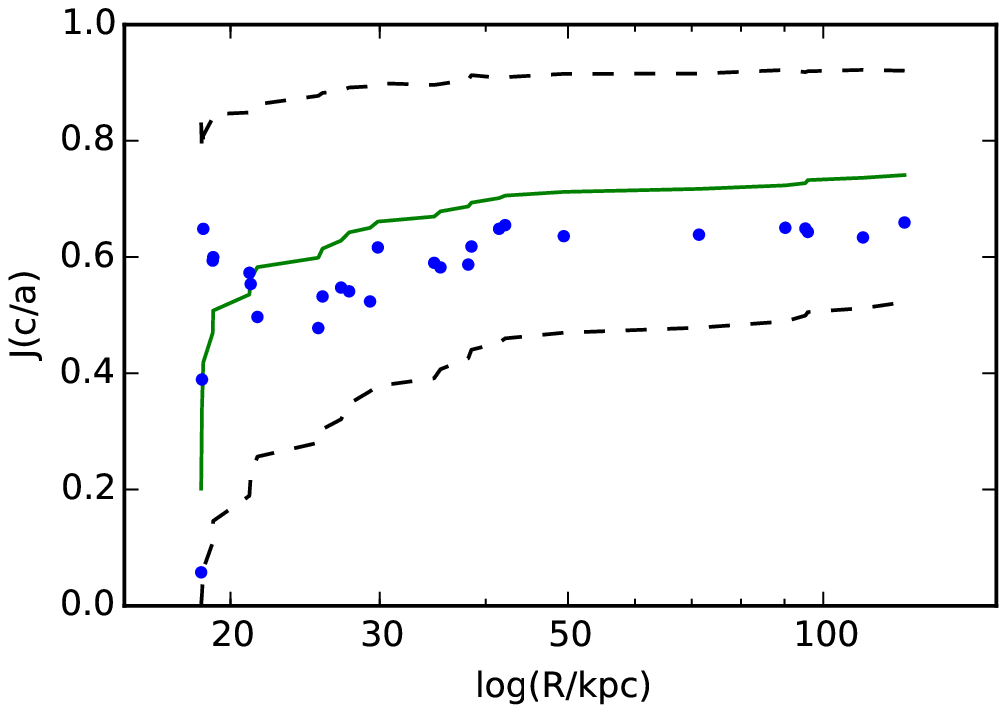}
 \includegraphics[width=0.325\textwidth]{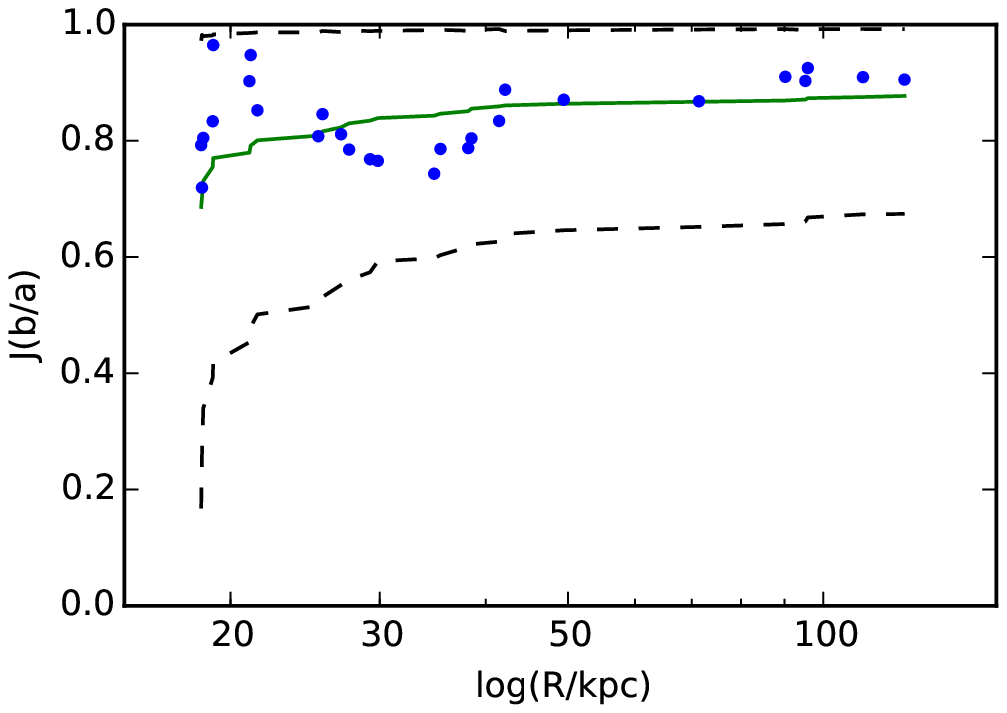}
 \includegraphics[width=0.325\textwidth]{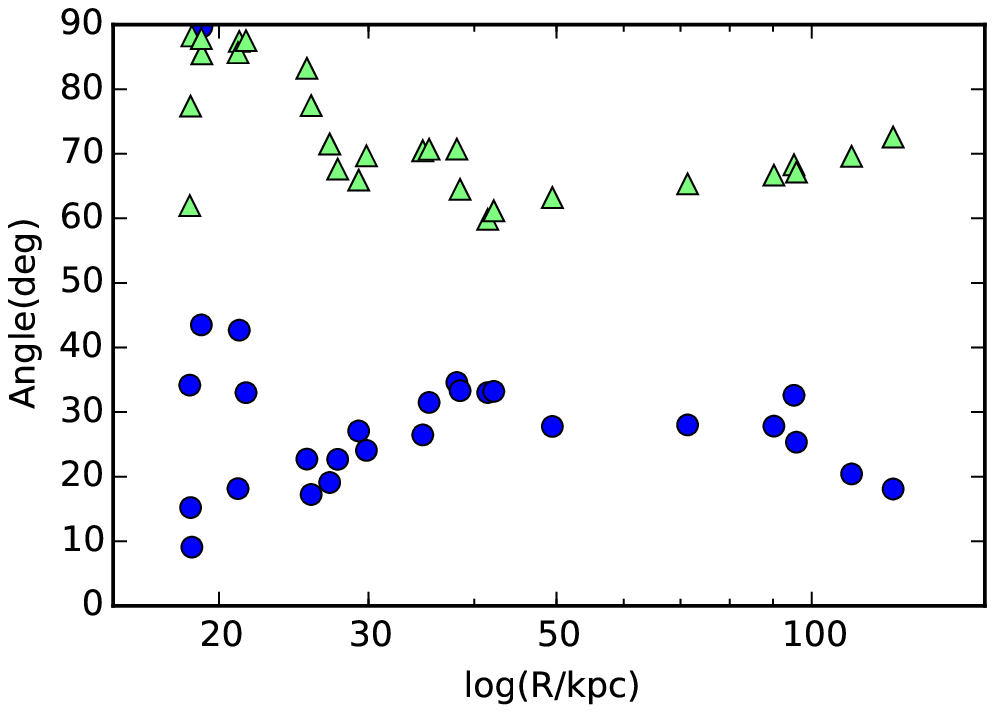} 
 \caption{Same as bottom panel of Fig.5, but for R>18 kpc for GCs . }
 \label{fig:7}
\end{figure*}

\subsection{Degree of anisotropy}
The anisotropic distribution of satellite galaxies around the Milky Way is a well known result, thus we start the current Section by demonstrating how the inertia tensor describes this well known anisotropy. The distribution of satellite galaxies by distance is demonstrated in Figure~\ref{fig:8}. We only test the satellite galaxies in the same region where we have GCs, i.e. the Milky Way, and hence we ignore more distant dwarfs belonging to the Local Group.
From the $c/a$ ratio shown in Figure~\ref{fig:9} for both tensors for the sample of 27 galaxies an explicit statistically significant anisotropy is observed at distances $R> 150$~kpc, but at smaller distances it is not very expressed. 
The similar situation is observed for a well known sample of 11 satellite galaxies, shown in Figure~\ref{fig:10}. The maximal deviation from the median of random samples is close to $3\sigma$ and is observed at the largest distances. The reduced tensor on the bottom panels shows somewhat lesser deviation from an isotropic distribution. In all cases the major axis of the distribution on the rightmost panels of Figures~\ref{fig:9}---\ref{fig:10} is directed almost perpendicular to the plane of the Galaxy. This demonstrates that our method finds the well known anisotropy, but it is less significant than with other methods \citep{2007MNRAS.374.1125M}.

From the $c/a$ ratio for the full GCs sample shown in the top left panel of Figure~\ref{fig:5} one can clearly see a significant anisotropy at $2\leq R \leq 10$~kpc. At $R>18$~kpc the $c/a$ ratio becomes close to the median for random isotropic samples. The $c/a$ ratio for the reduced tensor stays significantly lower than that for the isotropic case at $R>18$~kpc, what reflects the different weighting of the inertia and reduced tensors (128 out of 157 GCs are located at $R<18$~kpc).

We believe that the anisotropy at $R<18$~kpc is fully connected  with the Galactic disk. This is clearly seen from the rightmost panels of Figure~\ref{fig:5}: the major axis of the inertia tensor lies within the disk while the minor axis is perpendicular to the disk. The scale of 18~kpc is close to the radius of the Galactic disk which is found to be 14 kpc \citep{1992ApJ...400L..25R} or 25 kpc if the Monoceros ring is interpreted as a part of the disk \citep{2015ApJ...801..105X}.

To remove the influence of small-distance GCs on the reduced tensor (and to try to extract the anisotropy not connected with the disk), we repeat the analysis only for GCs with $R>18$~kpc. The results are shown in Figure~\ref{fig:7}. The distribution of $c/a$ shows no clear evidence of anisotropy.

The interpretation of the orientation of eigenvectors represented in the right panels of Figures~\ref{fig:5}-\ref{fig:7} for distant GCs at $R>18$~kpc is not so straightforward. The distribution of points in the top right panel of Figure~\ref{fig:5} at these distances seems to be chaotic, in agreement with the absence of anisotropy, and quite similar to the results for one random sample shown in Figure~\ref{fig:6}. The results in Figure~\ref{fig:7} are more stable and show that the distribution of GCs at $R>18$~kpc is slightly elongated in the direction perpendicular to the Galactic plane.
 
When analysing eigenvalue ratios $c/a$ and $b/a$ of the inertia tensors $S$ or $J$ defined in (\ref{form:1}-\ref{form:2}) as a function of distance one should note that 1) due to the limited statistics, even for 157 random points the median for random catalogues does not approach 1.0 closely; 2) the parameters we plot are correlated for points with close distances, which is clearly seen from Figure~\ref{fig:6}, especially for the angles measured using the reduced inertia tensor (lower right panel of Figure~\ref{fig:6}). This means that such a correlation itself cannot be used as an indication of some structure.

\begin{figure*}
 \includegraphics[scale=0.8]{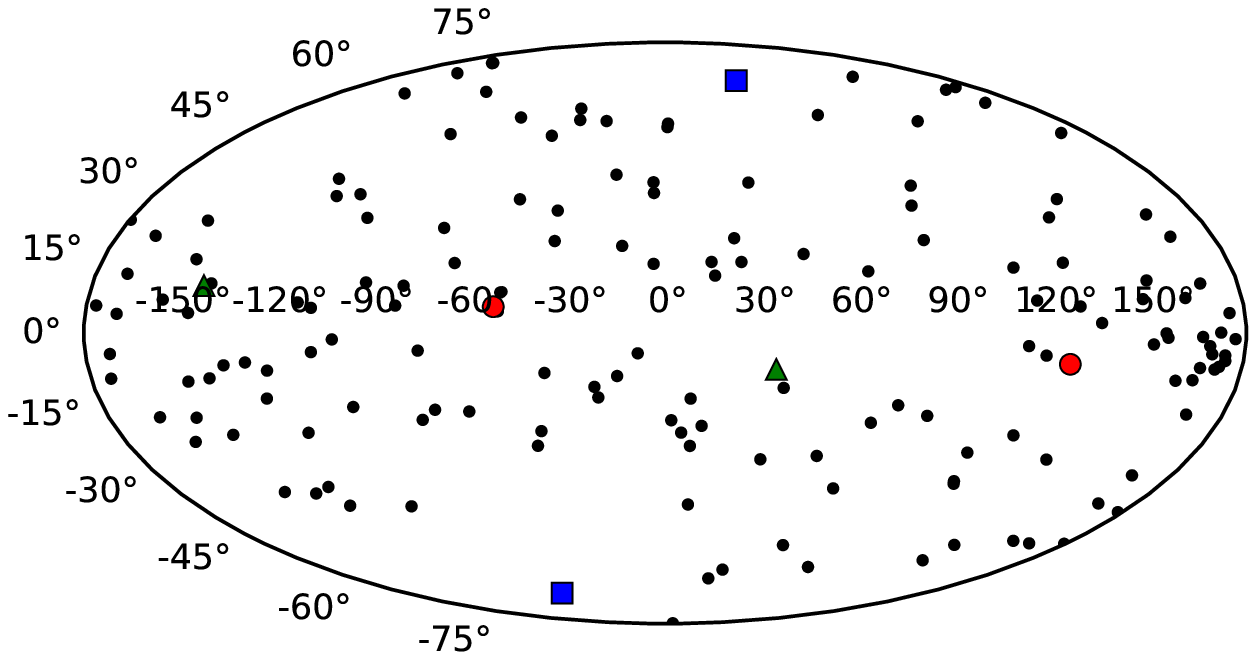}
 \caption{Map of the sky with the distribution of GCs in galactic coordinates. Square -- major axis of the Gyration tensor; Triangle -- intermediate axis of the Gyration tensor; Circle -- minor axis of the Gyration tensor.}
 \label{fig:14}
\end{figure*}

The distribution of the GCs on the sky together with the directions of the minor (circles), intermediate (triangles) and major (squares) axes for the full sample of GCs is shown in Figure~\ref{fig:14}. The observer is sitting in the galactic center.

\begin{figure*}
 \includegraphics[width=0.49\textwidth]{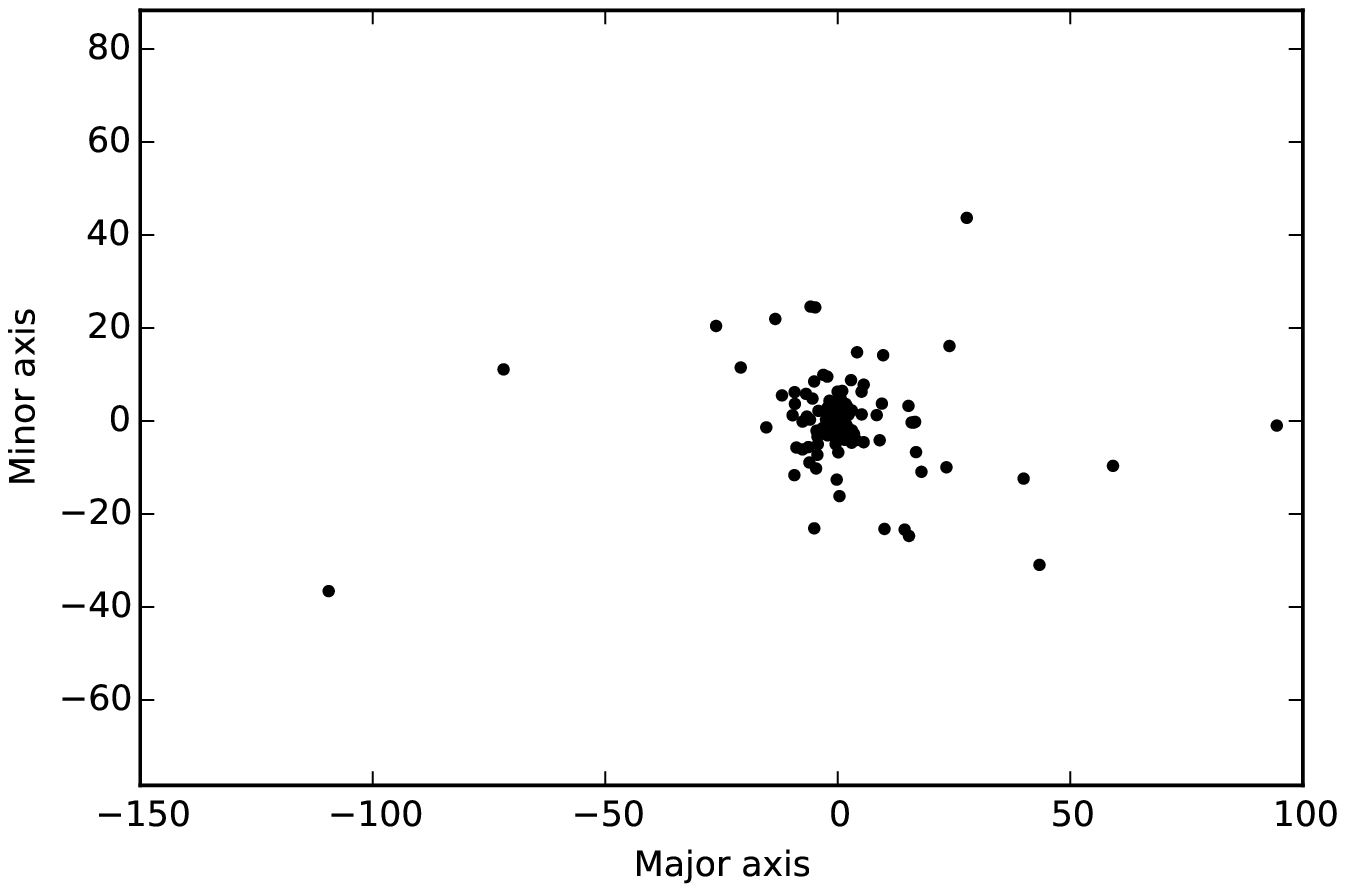} 
 \includegraphics[width=0.49\textwidth]{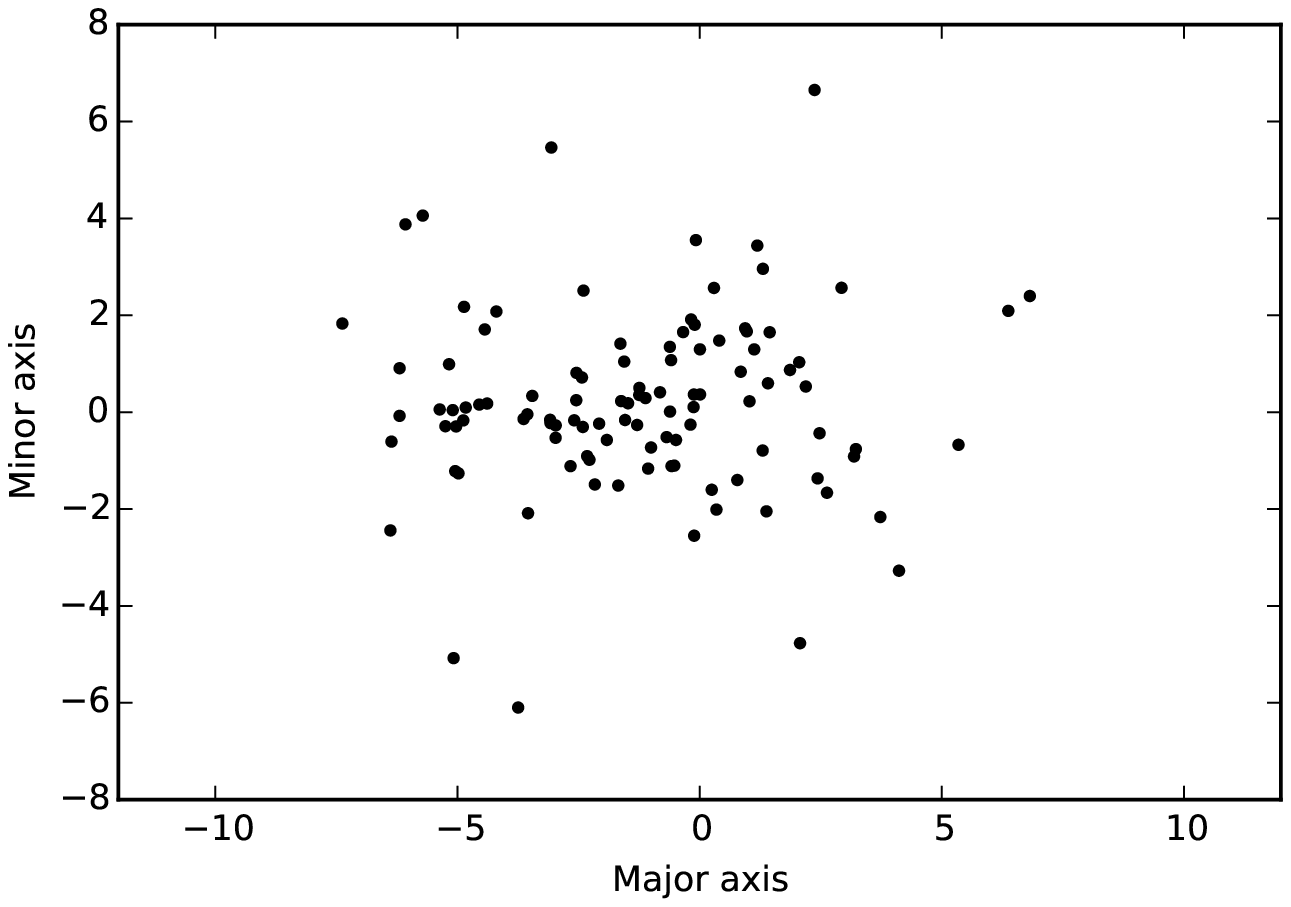} 
 \caption{Distribution of GCs along the largest and smallest eigenvectors. On the left panel for sampling R <125 kpc, and on the right panel for sampling R <10 kpc. }
 \label{fig:11}
\end{figure*}

\begin{figure*}
 \includegraphics[scale=0.8]{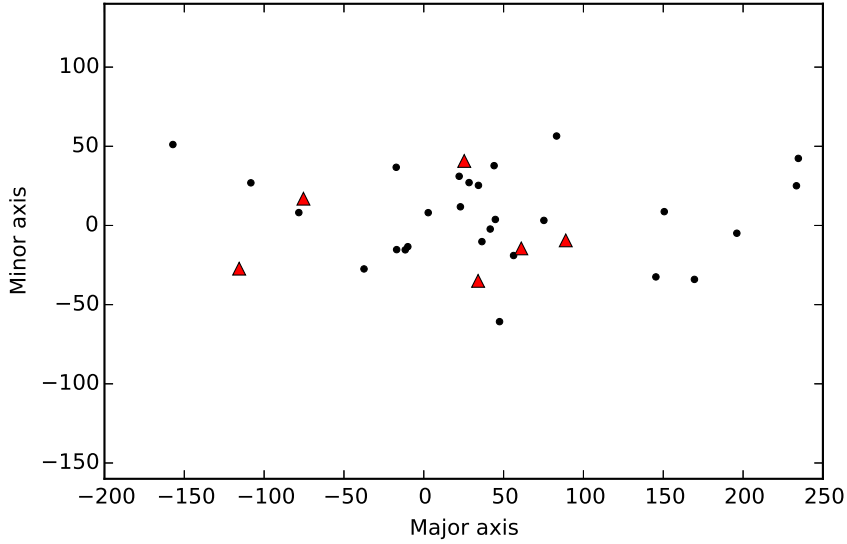}
 \caption{Distribution of satellite galaxies along the largest and smallest eigenvectors, with the addition of 6 most distant GCs. Circle -- satellite galaxies; Triangle -- GCs.}
 \label{fig:13}
\end{figure*}

The distribution of GCs along the largest and smallest eigenvectors is illustrated in Figure~\ref{fig:11}. The left panel shows the GCs distribution at a distance less than 125 kpc, which contains all 157 GCs. On the right panel, a distribution of 106 GCs that are located closer than 10 kpc to the Galactic center is shown. A disk-like structure is seen on the right panel by a naked eye.  In Figure~\ref{fig:13}, we show the distribution of the satellite galaxies along the smallest and largest eigenvectors. There, one can clearly distinguish an elongated structure.

\subsection{Comparison of distant GCs with the plane of satellites}
\label{sec:comp}

In Fig.~\ref{fig:13} one can see that the satellite galaxies form an elongated structure with a width of $\approx100$ kpc and a length of 400 kpc. Furthermore, for major axis $l=-132^\circ$, $b=73^\circ$ and for minor axis $l=154^\circ$, $b=-5^\circ$. It is interesting to check how the GCs are distributed in the same coordinate axes, defined by the eigenvectors of the system of satellite galaxies. We select only the GCs which lie on distances greater than 50 kpc (the half thickness of the plane of satellites) from the galaxy center, since GCs at smaller distances will lie entirely inside the structure formed by galaxies. There are only 6 GCs selected. In Figure~\ref{fig:13} we show their distribution together with that of satellites. One can see that all 6 GCs lie within the $\pm50$\,kpc region along the minor axis of satellite galaxies' inertia tensor. RMS half thickness for satellite galaxies is 28.7~kpc and for GCs is 26.5~kpc. This again shows the similarity of the GSs distribution at large distances with the distribution of satellite galaxies. 

To check if this finding is a result of coincidence, we again generate 10,000 random catalogues using our Method of anisotropy measurement, and find that the chance of probability of having all 6 GCs inside $\pm50$\,kpc is 1.7\%. This value does not allow to confidently conclude that the GCs are located on the same plane as satellites, however, it gives an indication that this may be true. Additional information is needed to answer the question if there is a real ``plane of GCs'' in our Galaxy. Analysis of GCs' proper motions should give this information.

\section{Anisotropy Measurement for Three Types of GCs}

The GCs were first divided into several types more than two decades ago \citep{1993ASPC...48...38Z,1993ApJ...411..178V}. The proposed classification was updated twice, by \citet{2004MNRAS.355..504M} and \citet{2005MNRAS.360..631M}. The classification of GCs was made on the basis of cluster metallicity and horizontal branch (HB) morphology. The metal--rich GCs which  have red HBs and are  confined  to  the  bulge  and  inner  disc  of  the  Galaxy are designated  ``bulge/disc'' (BD). While, the metal--poor  clusters with blue and red HBs  are so-called ``old halo'' (OH) and ``young halo'' (YH) ,  respectively and generally situated in the Galactic halo. In \citet{2005MNRAS.360..631M} classification GCs are divided into 5 types: GCs in the bulge / disk (BD -- 37 objects), GCs in the old halo (OH -- 70 objects); GCs in the young halo (YH -- 30 objects); GCs belonging to the tidal stream of Sagittarius (SG -- 6 objects); GCs of unknown type (UN -- 7 objects). From scenarios of GCs formation discussed in the Introduction it follows that different types of GCs should show different anisotropy. Here we focus on the analysis of anisotropy of only three most abundant types: BD, OH and YH. GCs in the BD lie at a distance from the center of the Galaxy from about 0.75~kpc to 17.5~kpc. In the OH and YH, GCs cover distances from 0.5~kpc to 90.2~kpc and from 1.4 to 125~kpc, respectively. 

The GCs are very concentrated towards the center in each of our samples: in the BD of the 37 GCs, only 8 GCs are located at distances greater than 5~kpc. In the OH of the 70 GCs, only 20 GCs are located at a distance more than 8~kpc. 
 
\begin{figure*}
 \includegraphics[width=0.325\textwidth]{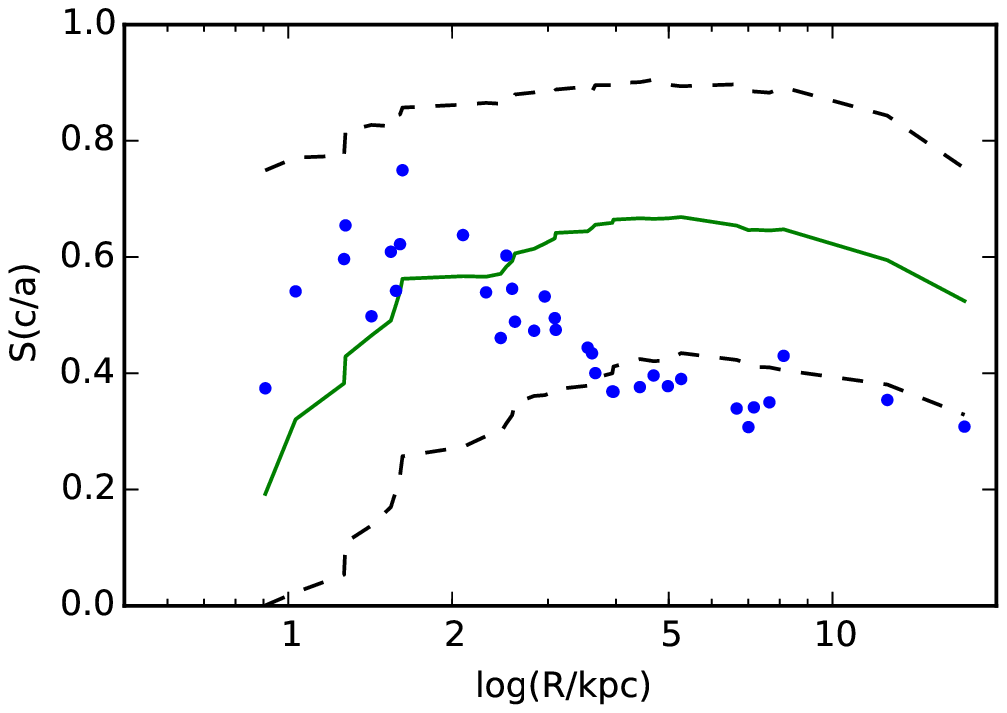}
 \includegraphics[width=0.325\textwidth]{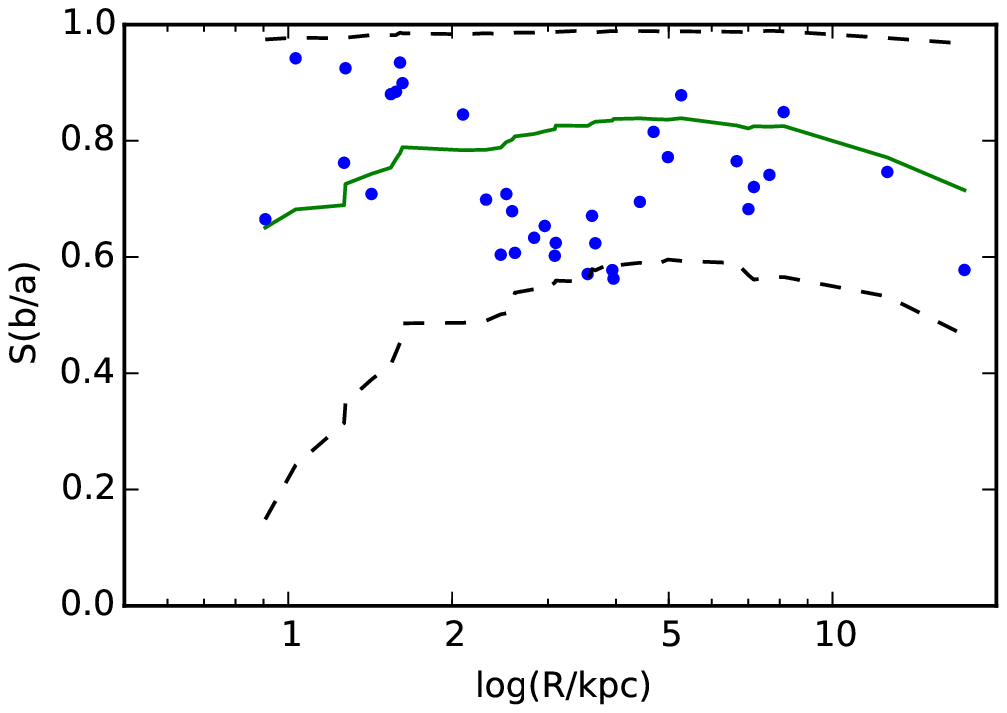}
 \includegraphics[width=0.325\textwidth]{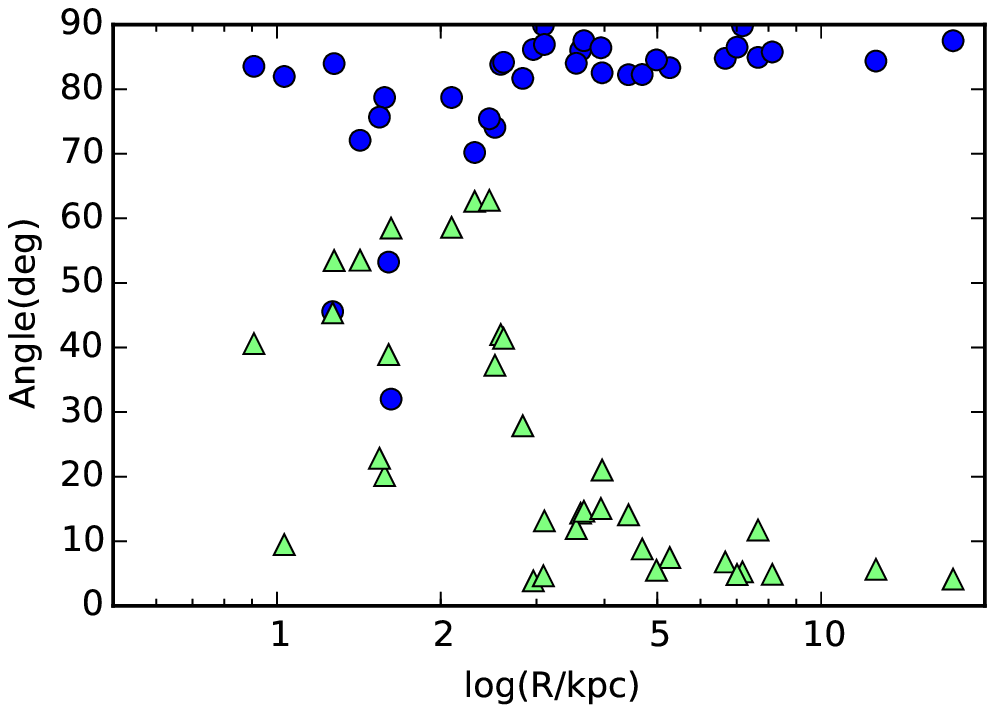} 
 
 \includegraphics[width=0.325\textwidth]{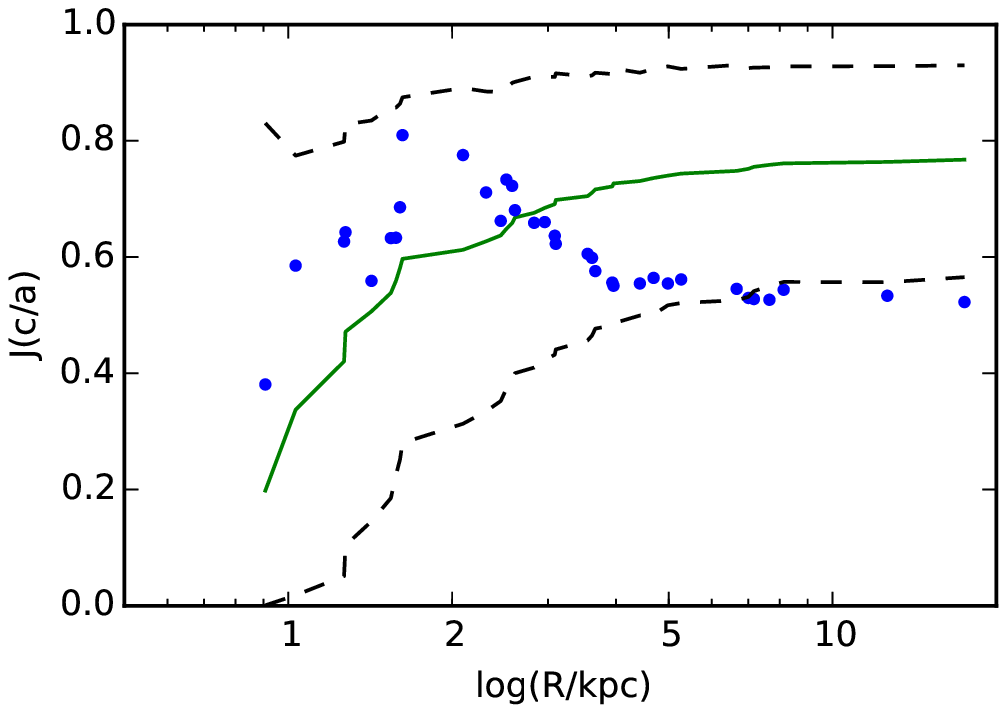}
 \includegraphics[width=0.325\textwidth]{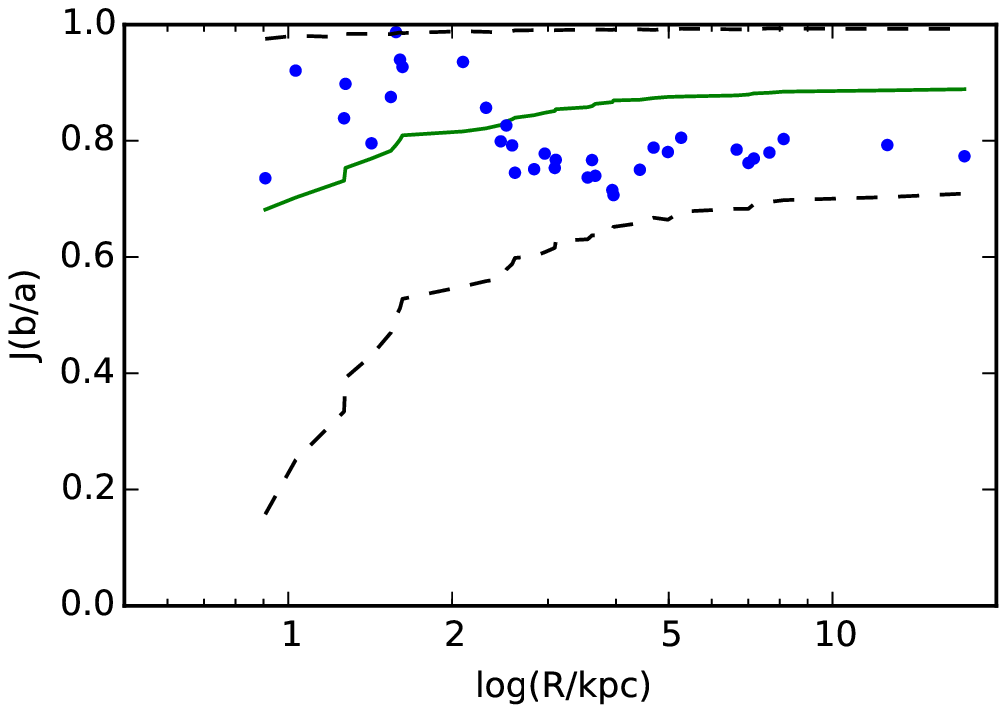}
 \includegraphics[width=0.325\textwidth]{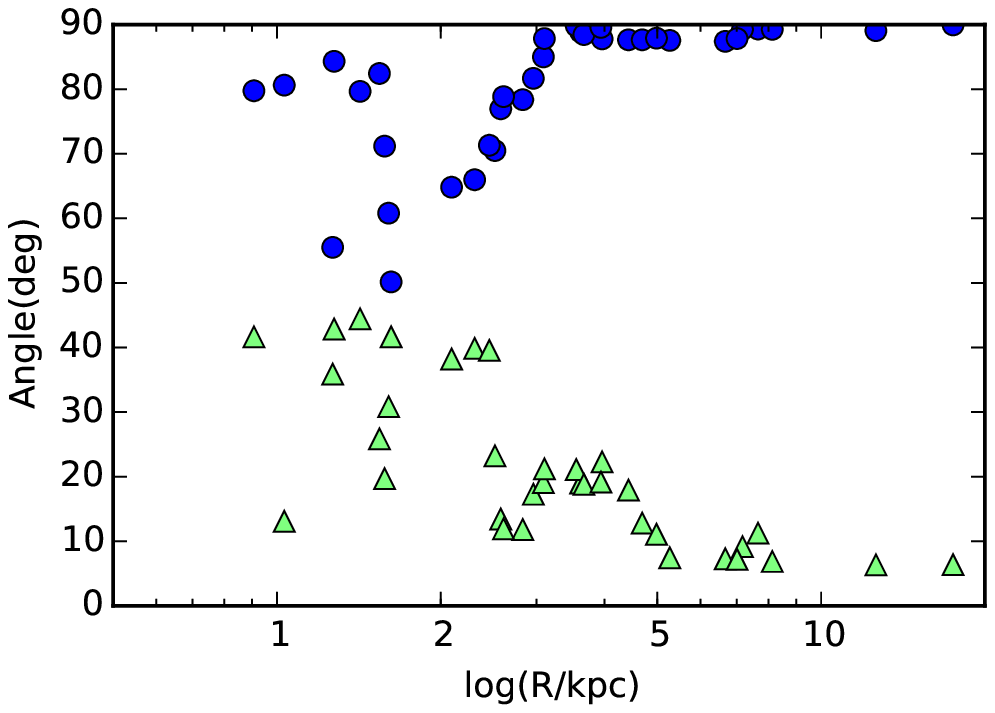}
 \caption{ Same as Fig.5, but for GCs in BD }
 \label{fig:17}
\end{figure*}

\begin{figure*}
 \includegraphics[width=0.325\textwidth]{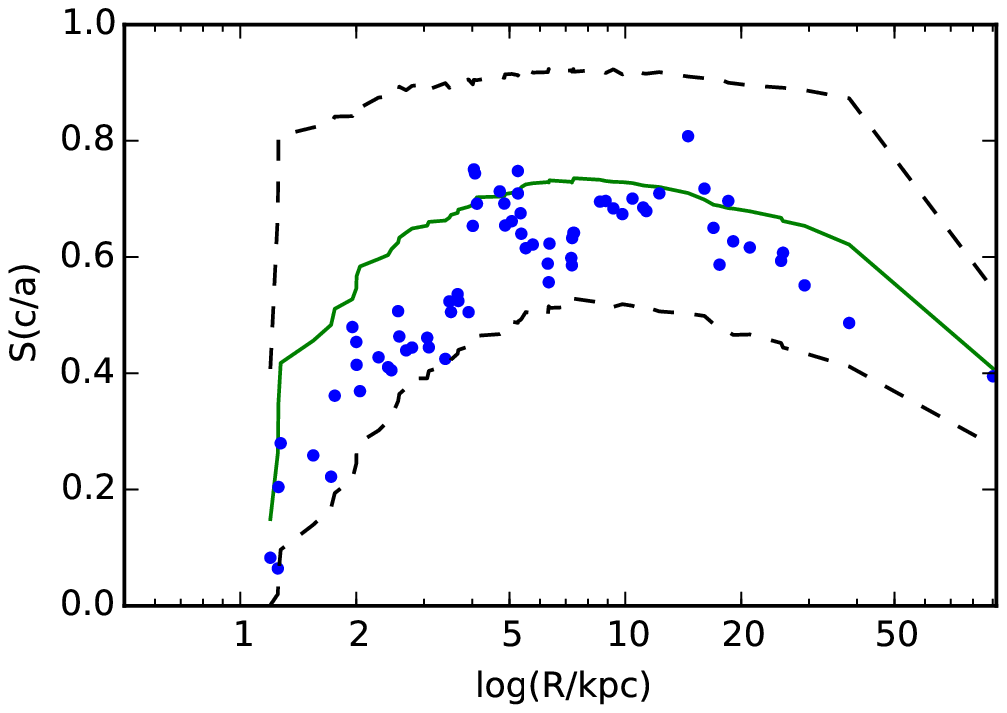}
 \includegraphics[width=0.325\textwidth]{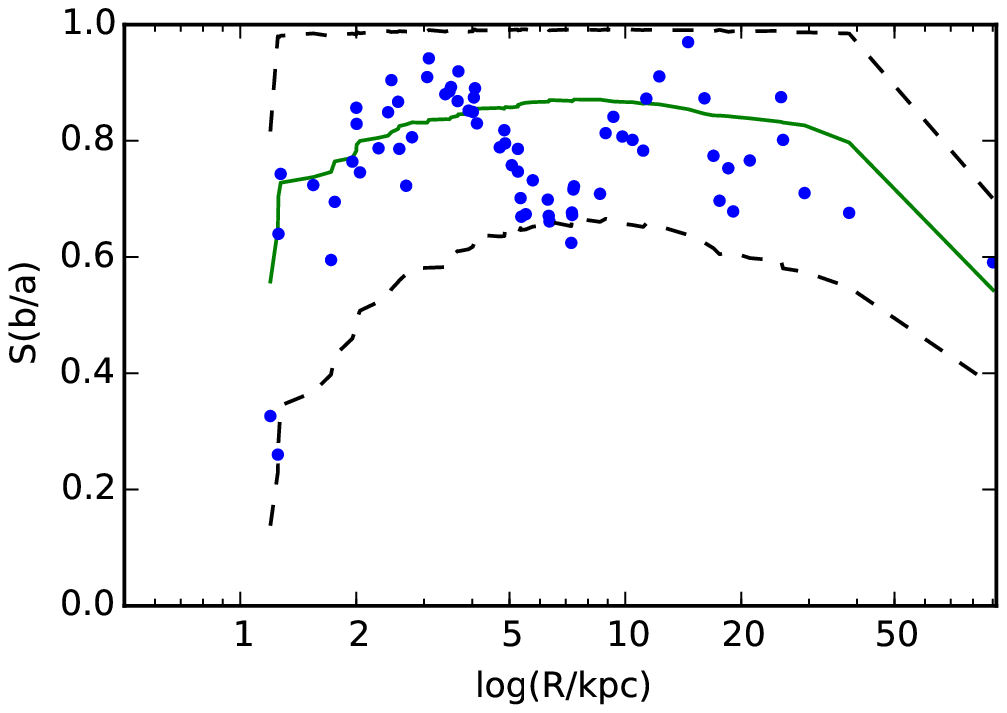}
 \includegraphics[width=0.325\textwidth]{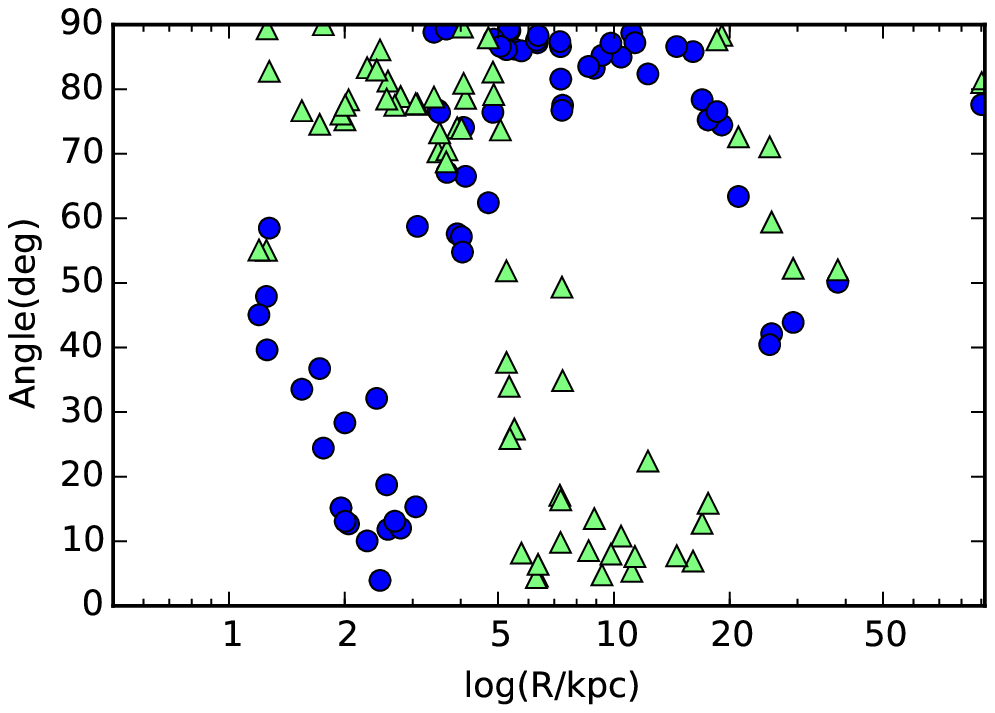} 
 
 \includegraphics[width=0.325\textwidth]{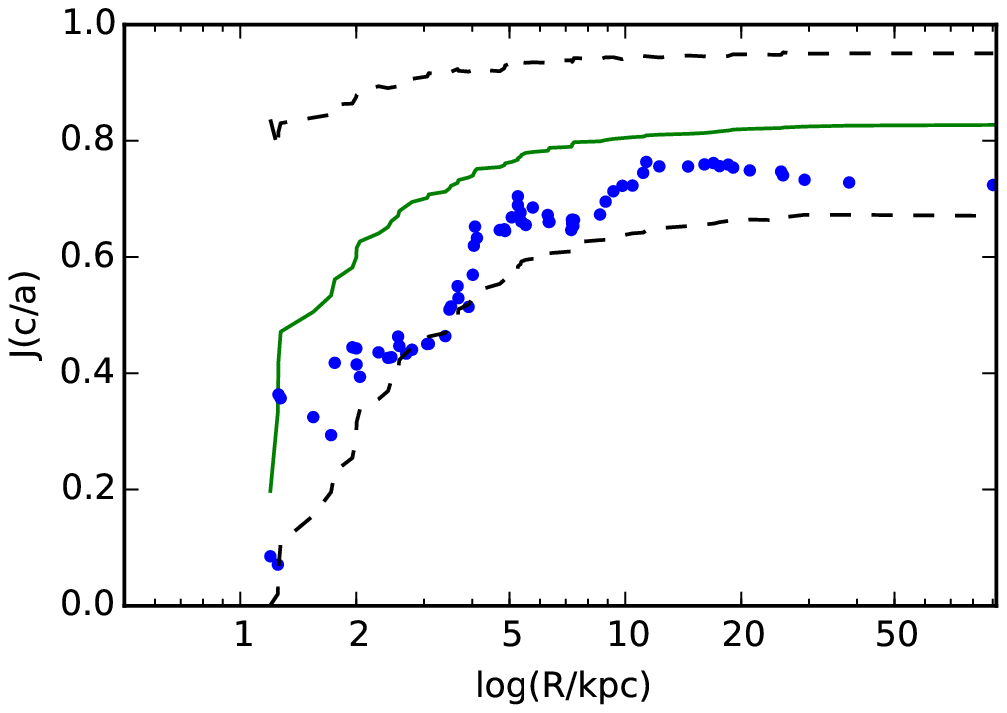}
 \includegraphics[width=0.325\textwidth]{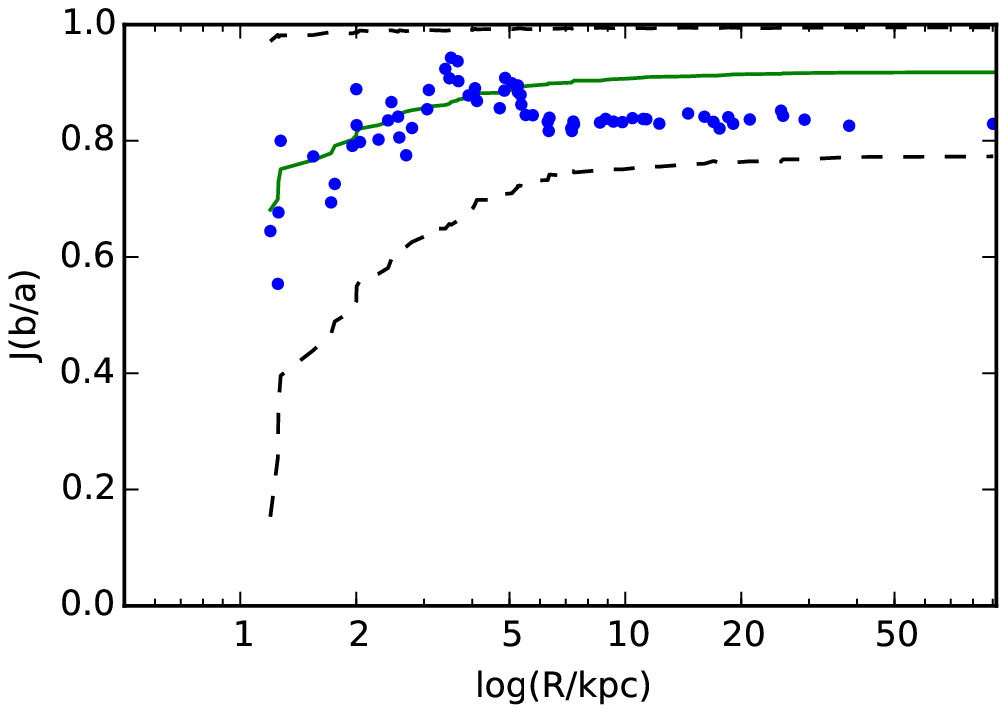}
 \includegraphics[width=0.325\textwidth]{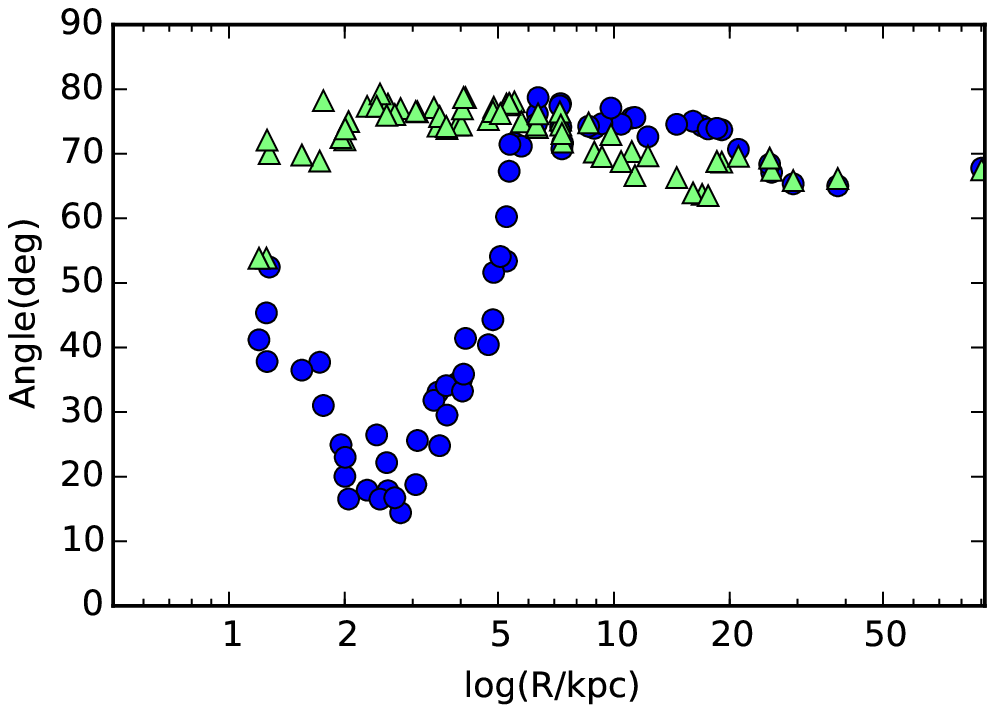}
 \caption{Same as Fig.5, but for GCs in OH }
 \label{fig:18}
\end{figure*}

For GCs in BD (Figure~\ref{fig:17}), the very central region with $R<2$~kpc has an isotropic distribution, while at $R>3.5$~kpc for $S$ and at $R>7$~kpc for $J$ tensor a statistically significant anisotropy is observed. The orientation of this anisotropic distribution coincides with the Galactic disc, which can be seen in the 3$^\mathrm{rd}$ column of Figure~\ref{fig:17}. For GCs in OH (Figure~\ref{fig:18}) the deviation from random isotropic samples is always below $3\sigma$, but it approaches it at $R\lesssim 3$~kpc. From the intermediate column of Figure~\ref{fig:18} it is seen that the deviation from isotropic case gradually increases in the range of distances $3<R<6$~kpc, then it again decreases by $R\sim 10$~kpc. At the same time on the rightmost panel of Figure~\ref{fig:18} one can see a change of orientation at the same range distances: at $R<3$~kpc the minor axis of the structure lies close the galactic plane, the major axis is close to the pole, while at $R\approx10$~kpc the situation is the opposite. We conclude that OH GCs represent two kinematically different samples: one is forming a polar structure close to the center of the Galaxy, and another one is distributed more like BD GCs.

\begin{figure*}
 \includegraphics[width=0.325\textwidth]{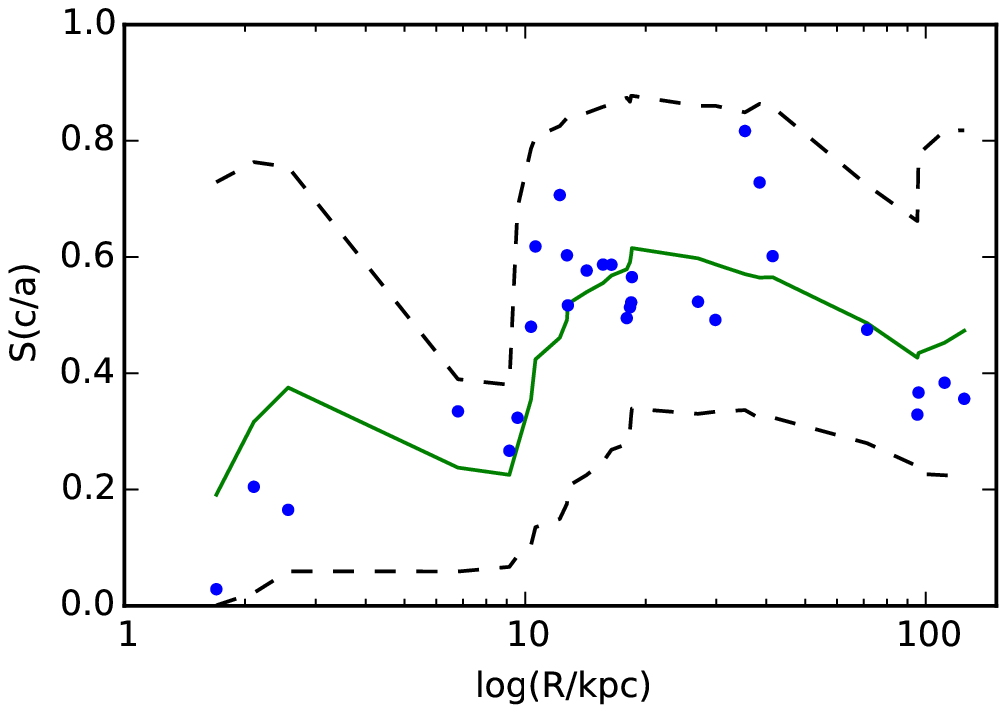}
 \includegraphics[width=0.325\textwidth]{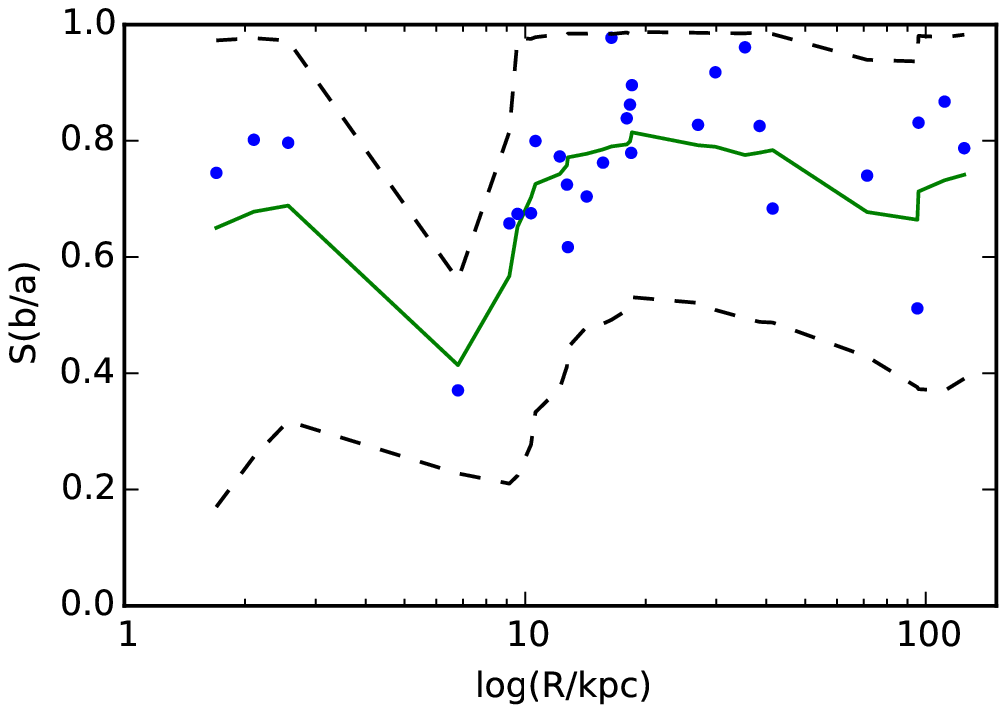}
 \includegraphics[width=0.325\textwidth]{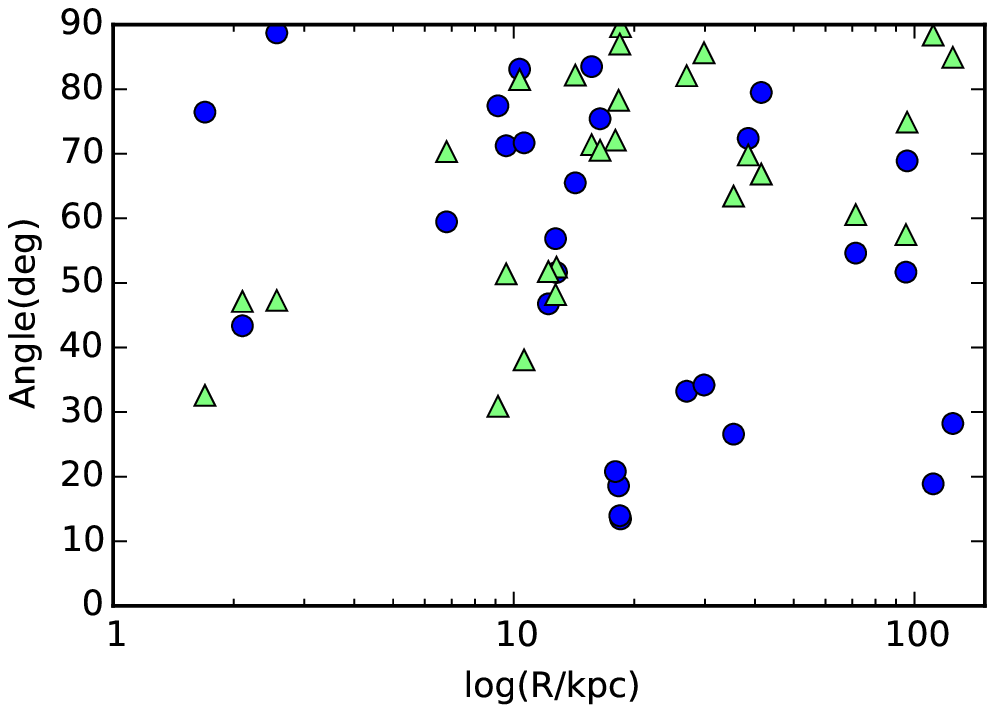} 
 
 \includegraphics[width=0.325\textwidth]{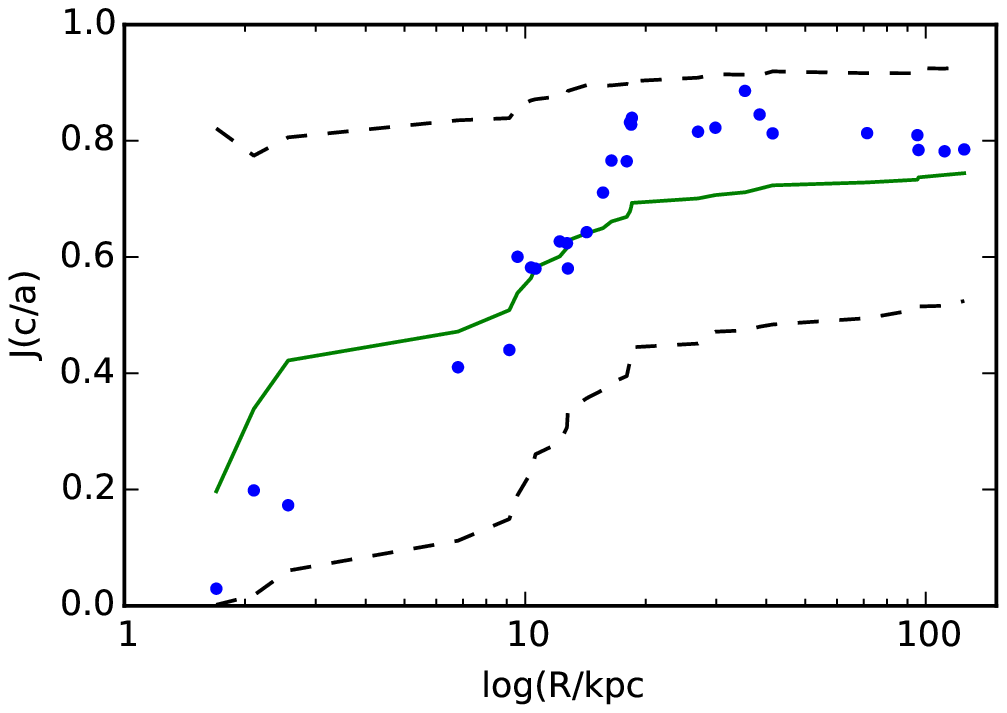}
 \includegraphics[width=0.325\textwidth]{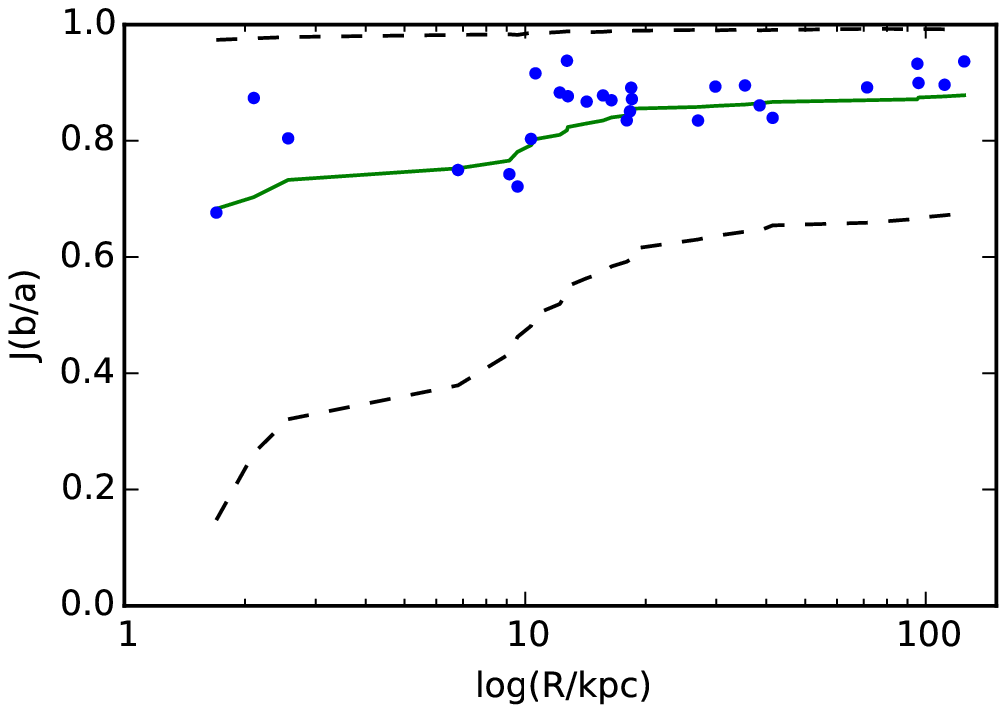}
 \includegraphics[width=0.325\textwidth]{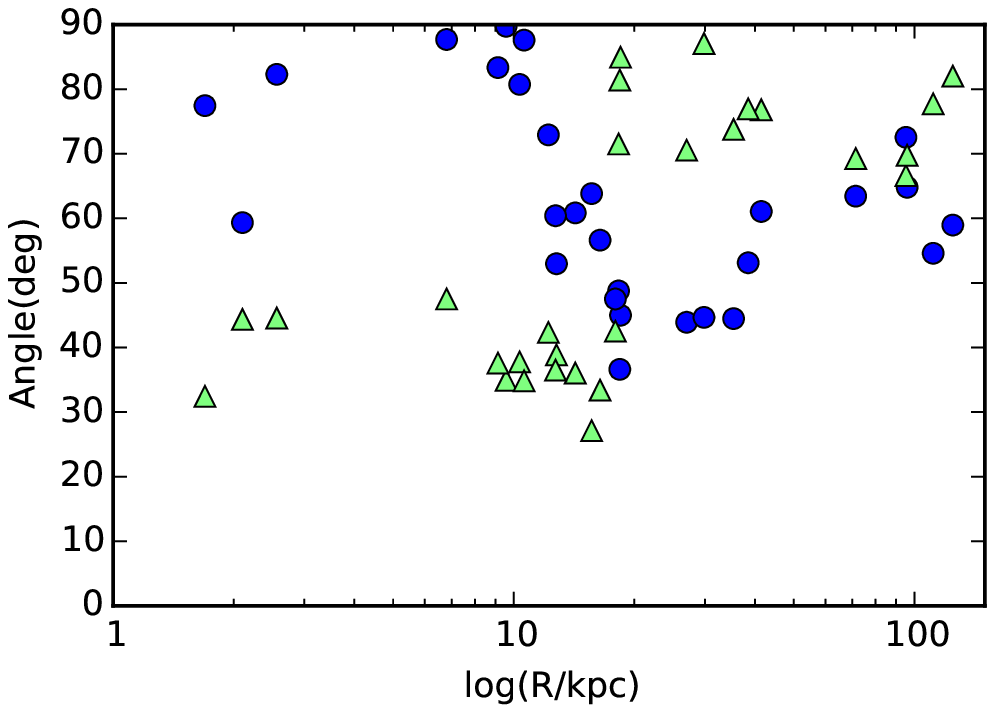}
 \caption{Same as Fig.5, but for GCs in YH }
 \label{fig:19}
\end{figure*}

The parameters of the YH GCs shown in Figure~\ref{fig:19} do not give any clear evidence of anisotropy, in contrast with the findings of \citet{2012ApJ...744...57K}. We interpret this discrepancy as due to a difference in the methods used and differences in the sample selection of YH GCs.

The third column in Figure~\ref{fig:19} shows that at two points at the largest distances the distribution of YH GSc is slightly elongated towards the galactic pole, as was shown in Section~\ref{sec:comp}.

The direction of the axes for all objects and in the regions where we found the anisotropy are presented in Table~\ref{tab:3}.

\begin{table*}
 \caption{Direction of axes for different samples}
 \label{tab:3} 
 \medskip
 \begin{tabular}{|l|c|c|c|c|}
  \hline
  \multicolumn{1}{|c|}{Sample} &  Major axis &  Major axis & Minor axis & Minor axis \\
    & l &  b &   l  &   b       \\
  \hline 
All GCs & 71$^\circ$ & 76$^\circ$ & 126$^\circ$ & $-8^\circ$ \\
  \hline
  GCs ($2<R<10$) & $-4^\circ$ & 2$^\circ$ & $-103^\circ$ & 80$^\circ$ \\    
  \hline
  All SGs $^{1)}$ & $-132^\circ$ & 73$^\circ$ & 154$^\circ$ & $-5^\circ$   \\
  \hline
  11 SGs $^{1)}$ & $-143^\circ$ & 64$^\circ$ & 153$^\circ$ & $-12^\circ$  \\
  \hline
  All BD & $-32^\circ$ & $-3^\circ$ & 21$^\circ$ & 86$^\circ$  \\
  \hline
  BD ($R>3$) & $-33^\circ$ & $-3^\circ$ & 20$^\circ$ & 85$^\circ$  \\
  \hline
  All OH & 3$^\circ$ & $-12^\circ$ & 91$^\circ$ & 9$^\circ$ \\
  \hline
  OH ($R<3$) & $-104^\circ$ & 78$^\circ$ & 99$^\circ$ & 11$^\circ$ \\
  \hline
  OH ($6<R<20$) & $-148^\circ$ & $-16^\circ$ & 122$^\circ$ & 1$^\circ$ \\
  \hline
  All YH & 62$^\circ$ & 62$^\circ$ & 142$^\circ$ & $-5^\circ$ \\   
  \hline     
  \end{tabular}
  \begin{tabular}{rp{150mm}}
 $^{1)}$  &  SGs -- satellite galaxies.  \\
\end{tabular}
\end{table*}

\section{Discussion and Conclusion}

Using the inertia tensor and the reduced tensor we characterize the anisotropy of GCs. We check that this method recovers the well known plane of satellites, however, its significance is somewhat lower than for other methods used in literature. 
For the GCs we find that the full sample shows significant anisotropy only in the range of distances $2<R<10$~kpc (see Figure~\ref{fig:5}). The structure has elongated shape with $c/a\approx0.5$ and $b/a\approx0.6$, with the major axis lying within the galactic plane. We believe that this structure is connected with the galactic disk. At distances $R<2$ and $R>18$ the parameters of the inertia tensor are very close to that for random isotropic samples.

Nevertheless, the spatial distribution of the 6 most distant GCs shows an indication of coincidence with the known planar structure in the distribution of satellite galaxies. The probability of random realization of such a distribution is 1.7\%. Measuring the proper motions of these GCs will shed light on their connection with the satellite galaxies: we expect that the proper motions will lie more or less within the indicated plane of satellites if the connection between GCs and satellites is real.

We also divide GCs into three classical types: BD, OH and YH and analyse their anisotropy separately. The BD GCs show isotropic distribution at $R<2$~kpc and a disk-like structure at $R>3.5$~kpc with $c/a\approx 0.3$ and $b/a\approx 0.6$ coplanar with the galactic disk (see Figure~\ref{fig:17}). The OH GCs show a more complex structure (see Figure~\ref{fig:18}). At small distances, $R<3$~kpc there is a cigar-like structure perpendicular to the galactic plane with $c/a\approx0.3$ and $b/a\approx 0.7$. At $R>6$~kpc it transforms  to an almost isotropic distribution, but slightly elongated with the major axis lying close to the galactic plane at $R<20$~kpc. We conclude that OH GCs represent two populations with dynamically different properties. The YH GCs (see Figure~\ref{fig:19}) do not show a clear anisotropy when analysed with the help of inertia tensor, but the major axis of the structure at highest distances is directed towards the galactic pole (see also Figure~\ref{fig:7}).

The scales of $R\approx2-6$~kpc that we found wherein the properties of the GCs distribution change are comparable with the scale of Galactic bulge --- 1.5 kpc and a bar --- 5 kpc \citep{2016ARA&A..54..529B}, while the noted $18$~kpc scale is close to the size of the disk, 14---25~kpc \citep{1992ApJ...400L..25R,2015ApJ...801..105X}.
Our findings indicate a complex distribution of the GCs at relatively small distances, $R<18$~kpc, which include two components of OH GCs with different anisotropy and a population of BD GCs coinciding with the galactic disk. They may reflect the anisotropic accretion satellites containing GCs, GCs formation inside the galactic disk, and the dynamical interaction with the disk and bulge gravitational potential. The analysis of the velocities of the GCs will shed more light on the properties of these populations. Another way to extend our analysis is to identify GCs belonging to  tidal streams, like the one of Sagittarius or Monoceros \citep{2014MNRAS.445.2971C,2005ApJ...626..128P} and explore the spatial distribution of GCs belonging and not belonging to the streams.

\section{Acknowledgements}
We thank Margarita Sharina for useful discussions and comments. 
The work is supported by the Program 28 of the fundamental research of the Presidium of the Russian Academy of Sciences ``Space: research of the fundamental processes and relations'', subprogram II ``Astrophysical objects as space laboratories''.

\bibliographystyle{mnras}
\bibliography{gc.bib}

\bsp	
\label{lastpage}
\end{document}